\documentclass[12pt,letterpaper]{JHEP3}

\usepackage{amsmath,amssymb,amsfonts,xspace}
\usepackage[latin1]{inputenc}
\usepackage{multirow}
\usepackage{epsfig}

\def\lfig#1#2#3#4{
\begin{figure}
\centerline{\hfill \includegraphics[height=#3]{#2}\hfill}
\caption{#1 \label{#4}}
\end{figure}
}

\numberwithin{equation}{section}
\def\Im{\,{\rm Im}\, }
\def\Re{\,{\rm Re}\, }

\def\rangl{\right\rangle   }
\def\langl{\left\langle  }
\def\({\left(}
\def\){\right)}
\def\[{\left[}
\def\]{\right]}
\def\hf{{1\over 2}}

\newcommand{\de}{\mathrm{d}}

\newcommand{\I}{\mathrm{i}}
\newcommand{\e}{\mathrm{e}}
\newcommand{\cL}{\mathcal{L}}

\def\vrh{\varrho}

\newcommand{\p}{\partial}

\newcommand{\half}{\frac{1}{2}}

\newcommand{\cU}{\mathcal{U}}

\newcommand{\cM}{\mathcal{M}}
\newcommand{\cN}{\mathcal{N}}

\newcommand{\vb}{\bar{v}}
\newcommand{\zb}{\bar{z}}
\newcommand{\wb}{\bar{w}}

\DeclareSymbolFont{AMSa}{U}{msa}{m}{n}
\DeclareSymbolFont{AMSb}{U}{msb}{m}{n}
\DeclareMathSymbol{\fieldR}{\mathalpha}{AMSb}{"52}

\newcommand{\N}{{\mathcal N}}
\newcommand{\kahler}{{K\"ahler}\xspace}
\newcommand{\hk}{{hyperk\"ahler}\xspace}

\newcommand{\cZ}{\mathcal{Z}}
\newcommand{\cO}{\mathcal{O}}

\renewcommand{\Im}{{\rm Im}}
\renewcommand{\Re}{{\rm Re}}

\newcommand{\pa}{\partial}
\newcommand{\nn}{\nonumber}

\newcommand{\eps}{\epsilon}
\newcommand{\IR}{\mathbb{R}}
\newcommand{\IC}{\mathbb{C}}
\newcommand{\IZ}{\mathbb{Z}}

\newcommand{\vth}{\vartheta}
\newcommand{\om}{\omega}
\newcommand{\Om}{\Omega}

\newcommand{\CP}{\IC P^1}
\def\bea{\begin{eqnarray}}
\def\eea{\end{eqnarray}}
\def\be{\begin{equation}}
\def\ee{\end{equation}}
\def\ba{\begin{align}}
\def\ea{\end{align}}
\def\bse{\begin{subequations}}
\def\ese{\end{subequations}}

\def\bU{ \bar U }

\def\bZ{\bar Z}

\def\ba{\bar a}

\def\bi{\bar \imath}
\def\bj{\bar \jmath}

\def\bw{\bar w}
\def\bx{\bar x}

\def\bz{\bar z}
\def\bu{\bar u}
\def\bv{\bar v}

\def\ze{\zeta}
\def\bze{\bar \zeta}

\newcommand{\CL}{{\cal{L}}}

\def\ztp{\zeta_+}
\def\ztm{\zeta_-}
\def\ztpm{\zeta_\pm}

\def\nupk#1{\hat\nu_{#1}}

\def\bnupk#1{\bar {\hat\nu}_{#1}}

\def\muor{\mui{0}}

\def\muinf{\mui{\infty}}

\def\hmu{\hat\mu}
\def\hnu{\hat\nu}

\def\ui#1{^{[#1]}}

\def\vi#1{v_{[#1]}}
\def\wi#1{w^{[#1]}}
\def\mui#1{\mu^{[#1]}}
\def\nui#1{\nu_{[#1]}}
\def\bvi#1{\bar v_{[#1]}}
\def\bwi#1{\bar w^{[#1]}}

\def\tmui#1{\tilde\mu^{[#1]}}
\def\tnui#1{\tilde\nu_{[#1]}}
\def\nuzi#1{\breve\nu_{[#1]}}
\def\muzi#1{\breve\mu^{[#1]}}
\def\mupi#1{\hat\mu^{[#1]}}
\def\nupi#1{\hat\nu_{[#1]}}
\def\rhoi#1{\rho^{[#1]}}

\def\cij#1{c^{[#1]}}
\def\Hij#1{H^{[#1]}}
\def\tHij#1{\tilde H^{[#1]}}
\def\Sij#1{S^{[#1]}}
\def\tSij#1{S^{[#1]}_T}
\def\Gi#1{G^{[#1]}}
\def\Ti#1{T^{[#1]}}
\def\Tiinv#1{{\tilde T}^{[#1]}}

\def\Hp{H_{\scriptscriptstyle\smash{(1)}}}
\def\tHp{{\tilde H}_{\scriptscriptstyle\smash{(1)}}}

\def\Hpij#1{\Hij{#1}_{\scriptscriptstyle{\smash{(1)}}}}
\def\tHpij#1{\tHij{#1}_{\scriptscriptstyle{\smash{(1)}}}}

\title{Linear perturbations of hyperk\"ahler metrics}

\preprint{PTA/08-045, ITP-UU-08-37,SPIN-08-28, \\IPhT-T08/104, LPTENS-08/34}

\author{Sergei Alexandrov$^1$, Boris Pioline$^{2}$,
Frank Saueressig$^3$, Stefan Vandoren$^4$\footnote{E-mail: 
{\tt alexandrov@lpta.univ-montp2.fr, pioline@lpthe.jussieu.fr,
frank.saueressig @cea.fr, S.J.G.Vandoren@uu.nl}}\\

$^1$ {\it Laboratoire de Physique Th\'eorique et
Astroparticules, CNRS UMR 5207, \\
Universit\'e Montpellier II, 34095 Montpellier Cedex 05, France}\\

$^2$ -- {\it Laboratoire de Physique Th\'eorique et Hautes
Energies, CNRS UMR 7589, \\
Universit\'e Pierre et Marie Curie,
4 place Jussieu, 75252 Paris cedex 05, France} \\
\, -- {\it Laboratoire de Physique Th\'eorique de l'Ecole Normale
Sup\'erieure, \\CNRS UMR 8549,
24 rue Lhomond, 75231 Paris cedex 05, France}\\

$^3$ {\it Institut de Physique Th\'eorique, CEA and CNRS  URA 2306, \\
 F-91191 Gif-sur-Yvette, France}\\

$^4$ {\it   Institute for Theoretical Physics and
           Spinoza Institute,
           Utrecht University,
           Leuvenlaan 4,
           3508 TD Utrecht,
           The Netherlands
           }
}


\abstract{We study general linear perturbations of a class of $4d$ real-dimensional
hyperk\"ahler manifolds obtainable by the (generalized) Legendre transform method.
Using twistor methods, we show that deformations can be encoded in
a set of holomorphic functions of $2d+1$ variables, as opposed to the functions
of $d+1$ variables controlling the unperturbed metric. Such deformations generically
break all tri-holomorphic isometries of the unperturbed metric. Geometrically,
these functions generate the symplectomorphisms which relate local complex
Darboux coordinate systems in different patches of the twistor space.  The deformed
K\"ahler potential follows from these data by a Penrose-type transform.
As an illustration of our general framework, we determine the leading exponential
deviation of the Atiyah-Hitchin manifold away from its negative mass Taub-NUT limit.
\\[1ex]
\vspace*{-5mm}}

\begin{document}

\section{Introduction}

Hyperk\"ahler (HK) and quaternionic-K\"ahler (QK) manifolds
appear in a variety of important situations in field and string theories,
yet their metrics are rarely known in closed form.
In this work, we provide a general formalism for describing
linear perturbations of a class of HK manifolds $\cM$ obtainable by
the (generalized) Legendre transform
method \cite{Karlhede:1984vr,Hitchin:1986ea,Lindstrom:1987ks,Ivanov:1995cy}. 
While our approach uses well-known twistorial techniques developed in
the seminal work  \cite{Hitchin:1986ea}, our explicit results offer a convenient
parametrization of deformations of HK metrics that should be useful
in many physics applications involving instanton corrections to moduli
space metrics. In the forthcoming 
work  \cite{Alexandrov:2008nk}, we shall extend our methods to describe 
linear perturbations of QK manifolds related to the above ones
by the superconformal quotient construction \cite{MR1096180,deWit:2001dj}, 
and in  \cite{apsv3} we shall use these results to determine
the general form of D-instanton corrections to the hypermultiplet  
moduli space in type II string theory compactifications

Before turning to geometry, it is worthwhile elaborating on physics motivations.
HK (resp. QK) manifolds generally arise as the target
space of rigidly (resp. locally) supersymmetric sigma models with
eight supercharges~\cite{AlvarezGaume:1980vs,Bagger:1983tt}.
Examples of HK manifolds include the classical moduli space of
instantons~\cite{atiyah1978ci}, magnetic monopoles~\cite{Atiyah:1988jp} and
Higgs bundles~\cite{hitchin1987sde} in four, three- and two-dimensional Yang-Mills
theories respectively, the quantum moduli
space of $\cN=4$ gauge theories in three dimensions~\cite{Seiberg:1996nz}, or
$K3$ surfaces that appear as part of the target space
for type II (resp. heterotic) superstrings with 16 (resp. 8) supercharges.
On the other hand, QK manifolds describe the
hypermultiplet moduli spaces of type II superstrings compactified on a
Calabi-Yau three-fold $X$~\cite{Cecotti:1989qn}, or heterotic strings
compactified on $K3$ \cite{Aspinwall:1998bw,Witten:1999fq}. This sector
of string theory, far less understood than its vector multiplet cousin (easily described
by special K\"ahler geometry), is the prime motivation for the present work.

In many cases, the geometry is well understood near
asymptotic infinity, where the underlying physics admits a weakly coupled
description. This is the case for the monopole moduli space
at large monopole separations, where the
metric asymptotes to a higher-dimensional version of negative-mass
Taub-NUT space \cite{Gibbons:1995yw}, and
also for the moduli space of three-dimensional $\cN=4$ gauge
theories, where the perturbative series truncates at one-loop -- in fact,
the two moduli spaces are conjecturally identical \cite{Seiberg:1996nz}.
In type II string compactifications, the tree-level asymptotic
hypermultiplet metric
may be obtained from the metric on the vector multiplet moduli space
(exactly computable via mirror symmetry) by the $c$-map
construction~\cite{Cecotti:1989qn,Ferrara:1989ik}.
The one-loop correction is determined by the Euler characteristic of $X$,
and there are indications that there are no perturbative
corrections beyond one loop~\cite{Robles-Llana:2006ez}.
On the heterotic side, the hypermultiplet
metric is in principle exactly computable in conformal field theory, but is usually
known only in the weak curvature regime \cite{Witten:1999fq}. In all these
cases, the asymptotic geometry is toric, i.e. 
possesses the required number of commuting isometries to be obtainable from 
the Legendre transform construction \cite{Karlhede:1984vr,Hitchin:1986ea}.
We return to the geometric significance of this class of metrics below.

Away from this weak coupling region, the metric is usually poorly understood,
or in some cases known only implicitly. 
A notable exception
is the moduli space of two $SU(2)$ BPS monopoles, the Atiyah-Hitchin (AH) manifold,
whose $SU(2)$-invariant metric may be expressed in terms of elliptic
functions~\cite{Atiyah:1988jp}, and obtained by the generalized Legendre transform
construction~\cite{Lindstrom:1987ks} from an $\cO(4)$ projective multiplet~\cite{Ivanov:1995cy}.
In the limit of large monopole separation, the $\cO(4)$ multiplet degenerates into the
square of an $\cO(2)$ multiplet, and the AH manifold
reduces to negative-mass Taub-NUT  \cite{Gibbons:1995yw}. At finite separation
however, exponential corrections break all tri-holomorphic isometries (see e.g.
\cite{Hanany:2000fw}), although a tri-holomorphic higher rank Killing tensor
does remain \cite{bielawski2000tqh,dunajski2003tth}. More generally,
a description of the moduli space of $k$ $SU(2)$ monopoles using
$\cO(2j)$ $(1\leq j\leq k-1)$ projective supermultiplets is known~\cite{Houghton:1999hr},
but only in a rather implicit form. Determining the leading exponential correction at long
distance is one of the possible applications of the
techniques developed in this paper.

Physically, corrections to the asymptotic metric correspond to instanton
contributions in field or string theory. In quantum field theory, there is
a well-defined procedure to compute these corrections in the one- or
two-instanton approximation.
Using these techniques, the conjectural equality between
the monopole and the 3D field theory moduli spaces was verified in
a number of cases \cite{Dorey:1997ij,Fraser:1997xi,Dorey:1998kq}.
In string theory however, the rules of space-time instanton calculus have not
been derived from a microscopic Lagrangian, although they may sometimes be inferred
indirectly. 

For type IIA string theory (or M-theory) compactified on a Calabi-Yau three-fold,
the general form of instanton corrections to the hypermultiplet
moduli space was first investigated in \cite{Becker:1995kb}.
More recently, the S-duality  of type IIB string theory 
was used to derive instanton corrections to the hypermultiplet moduli space
in general type II compactifications on a Calabi-Yau $X$,
for a subset of instanton configurations which preserve
the toric isometries~\cite{RoblesLlana:2006is,
RoblesLlana:2007ae}: namely  $D(-1)$, $F1$ and $D1$
instantons in type IIB, or  $D2$-branes in an appropriate Lagrangian
sublattice in $H^3(X,\IZ)$ (i.e.  ``wrapping A-cycles only'', for a suitable choice of
symplectic basis of A and B-cycles) in type IIA. Agreement with the analysis
of \cite{Ooguri:1996wj} in the conifold limit was demonstrated
in \cite{Saueressig:2007dr}.

On the other hand, it has remained an open problem to include the $D3$
and $D5$ instantons in type IIB (or the D2-branes wrapping B-cycles
in type IIA), as well as $NS5$ instantons in either theory. As the K\"ahler classes
of $X$ (in type IIA, the complex structure) are taken to infinity, these
instanton effects are much smaller than the ones already taken into
account. Thus, it is natural to treat them as linear perturbations away from
the metric found in \cite{RoblesLlana:2006is}. This was carried out
for the ``universal hypermultiplet'' in \cite{Alexandrov:2006hx} using 
techniques germane to four-dimensional QK manifolds, and our
aim here and in \cite{Alexandrov:2008nk} is to develop
similar techniques valid in any dimension.

As indicated above, the relevant $4d$-dimensional HK metrics in the weak coupling region are
obtainable by the generalized Legendre transform method. This construction, 
first uncovered in the physics literature using projective 
superspace techniques~\cite{Karlhede:1984vr,Lindstrom:1987ks}, was interpreted 
mathematically in the language of twistors in \cite{Hitchin:1986ea,Ivanov:1995cy,
bielawski2000tqh,Lindstrom:2008gs}\footnote{We thank M. Ro\v{c}ek and U. Lindstr{\"o}m 
for informing us
of their upcoming work \cite{Lindstrom:2008gs}.}. In particular, is was shown in  \cite{bielawski2000tqh}
that such metrics   -- which we refer 
to as $\cO(2n)$ HK spaces for brevity -- are characterized
by the existence of a local Hamiltonian action of a $d$-dimensional abelian  $\cO(2-2n)$-twisted
group. They can be encoded in a set of 
holomorphic functions $\Hij{ij}(\nu^I,\zeta)$ of  $d+1$
variables on the twistor space $\cZ$ of $\cM$  (one
such function for each pair of patches $(ij)$ needed to cover $\cZ$, subject to 
consistency conditions on the overlap of three patches, to local
gauge transformations and reality conditions).
The case $n=1$ corresponds to ``toric" HK manifolds with $d$ commuting tri-holomorphic isometries, obtainable by the standard Legendre transform
construction \cite{Hitchin:1986ea},
based on $\cO(2)$ projective supermultiplets. Cases with $n>1$ correspond
to HK geometries (such as the Atiyah-Hitchin manifold) with a higher rank 
Killing tensor, obtainable by the
generalized Legendre transform method based on $\cO(2n)$ projective
supermultiplets \cite{Lindstrom:1987ks}.

In this  work, we provide a framework to describe general linear perturbations
of  $4d$ real-dimensional HK manifolds $\cM$ obtainable by the 
generalized Legendre transform, with special emphasis on $\cO(2)$ manifolds.
By studying the deformations
of the holomorphic symplectic structure on the twistor space,
we show that deformations of $\cM$ preserving the HK property
may be encoded in holomorphic functions
$\Hpij{ij}(\nu^I,\mu_I,\zeta)$ ($I=1\dots d$) of $2d+1$ variables. Such perturbations
generically break all isometries (or Killing tensors) of the unperturbed geometry.
For $\cO(2)$
metrics,  we provide explicit formulae \eqref{mui}, \eqref{nupi} for the
complex coordinates of the perturbed
metric, and we show that the deformation of the K\"ahler potential \eqref{kahlerdef} in any
complex structure can be written as a Penrose-type contour integral
along the fibers of the projection $\pi: \cZ\to \cM$ (in particular, it is a
zero-eigenmode of the Laplace-Beltrami operator, as required by the
linearization of the Monge-Amp\`ere equation).
Deformations of $\cO(2n)$ metrics could be described by a straightforward 
generalization of our methods.
In the process, we also clarify the significance and ambiguity
of the contours arising in the Legendre transform construction
of $\cO(2n)$ HK manifolds.
We do not address possible obstructions at the non-linear level, nor 
whether these perturbations lead to geodesically complete metrics.

The outline of this paper is as follows.
In Section 2, we review the relation between the HK geometry of $\cM$ and the
holomorphic symplectic geometry of its twistor space $\cZ$,  and present a
general construction of $\cZ$ by patching together local Darboux complex
coordinate systems, using complex symplectomorphisms as transition functions.
In Section 2.4, we parenthetically argue that this provides a natural framework for the holomorphic
quantization of $\cZ$, with possible applications to topological string theory.
In Section 3, we specialize this construction to the case of $\cO(2n)$ manifolds,
and show how the standard generalized Legendre transform construction is
recovered. In particular, we obtain an important general formula \eqref{mui} for the twistor
lines in the $\cO(2)$ case.
In Section 4, we study general linear perturbations of $\cO(2)$ manifolds. 
Finally, in Section 5 we illustrate our method on the Taub-NUT and
Atiyah-Hitchin manifolds. In particular, we give a simple and elegant representation, Eq. \eqref{HtAH},
of the  leading exponential deviation of the AH manifold away from its Taub-NUT limit. 
In Appendix A, we show by a direct
computation that the deformed geometry is still \hk.

\section{General framework}

\subsection{Twistorial construction of general \hk manifolds\label{sec-rev}}

We start with a brief review of the twistor approach to \hk manifolds
(see \cite{MR664330,quatman,Hitchin:1986ea} for more details).
Let $\cM$ be a $4d$ real-dimensional \hk space. $\cM$ admits
three integrable complex structures $J^i$ ($i=1,2,3$) satisfying
the quaternion algebra
\be
J^i\, J^j = \eps^{ijk} J^k - \delta^{ij}\ ,
\ee
where $\eps_{ijk}$ is the totally antisymmetric tensor
with $\eps_{123}=1$ and $\delta_{ij}$ is the Kronecker
delta symbol. The metric $g(X,Y)$ is hermitian with respect
to each of the three complex structures, and each of the \kahler forms
\be
\omega^i(X,Y)=g(J^i X, Y)= -\omega^i(Y,X)\ ,
\ee
is closed. With respect to the complex structure $J^3$,
the two-forms $\omega^+=-\frac12(\omega^1 -\I\omega^2)$
and $\omega^-=-\frac12(\omega^1+\I\omega^2)$ are holomorphic 
and anti-holomorphic, respectively. More generally, any
 linear combination
\be
J(\ze,\bze)=\frac{1-\ze\bze}{1+\ze\bze} J^3
+\frac{\ze+\bze}{1+\ze\bze} J^2
+ \I \frac{\ze-\bze}{1+\ze\bze} J^1\ ,
\ee
where $\zeta \in \IC \cup \infty=\CP$ (and $\bze$ its complex conjugate) defines a complex structure
on $\cM$ compatible with the metric $g$. The corresponding
\kahler form is
\be
\omega(\ze,\bze)=\frac{1}{1+\ze\bze}
\left[(1-\ze\bze) \om^3 - 2\I\ze \om^+ + 2\I \bze \om^-  \right]\ ,
\ee
while
\be
\Omega(\zeta) = \omega^+ -\I \zeta\, \omega^3 + \zeta^2\, \omega^- \ ,
\ee
is a (2,0) form with respect to the complex structure $J(\ze,\bze)$ 
\cite{Hitchin:1986ea}.
The product $\cZ=\cM \times \CP$, equipped with
the complex structure $J(\ze,\bze)$ on $\cM$
times the standard complex structure
on $ \CP$, is a holomorphic fiber bundle on  $\cM$ known
as the twistor space of $\cM$. It is a trivial fiber bundle topologically,
but not holomorphically. We denote by $p: \cZ\to \CP$ the projection
$(m,\ze) \mapsto \ze$, and by $T_F={\rm Ker}({\rm d}p)$ the tangent bundle along
the fibers of this projection. Fixing a point $m$ on the base, the
set $\{m\} \times \CP$ is a holomorphic section of $p$, known
as a twistor line.

The parameter $\zeta$ may be viewed as a local complex
coordinate around the north pole $\zeta=0$ of the projective line $\CP$,
and $\Omega$ is regular around the north pole (in fact for all finite $\zeta$).
To remind of this fact, we write $\zeta\equiv\zeta\ui{0}$,
$\Omega(\zeta)\equiv\Omega\ui{0}(\zeta\ui{0})$.
At the south pole $\zeta=\infty$ however,  $\Omega(\zeta)$ has
a second order pole. It is convenient to reabsorb this divergence by defining
\be
\label{Ominfty}
\Omega\ui{\infty}(\zeta\ui{\infty})\equiv \zeta^{-2}\, \Omega(\zeta)
= \omega^- -\I\ze\ui{\infty} \omega^3
+ (\ze\ui{\infty})^2 \omega^+\ ,
\ee
where $\zeta\ui{\infty}=1/\ze\ui{0}$ is a local coordinate around the south pole.
The new $\Omega\ui{\infty}(\zeta\ui{\infty})$ is still a holomorphic two-form
in $J(\ze,\bze)$, but it is now regular around the south pole $\zeta\ui{\infty}=0$.
Note that the prefactor $ \zeta^{-2}$ in \eqref{Ominfty} is the second power
of the transition function $f_{\infty 0} =1/\zeta$ of the line bundle $\cO(1)$
over $\CP$ (see Section 2.2). Thus, $\Omega\ui{0}$ and
$\Omega\ui{\infty}$ should be viewed as  local trivializations
of a global section $\Omega$ of the holomorphic vector bundle
$\Lambda^2 T_F^*(2)$ \cite{Hitchin:1986ea}.

While the procedure just described produces a holomorphic
form $\Omega^{[\infty]}$ regular at the south pole, an a priori
different procedure is to apply the antipodal map
\be
\tau: \, (m,\ze,\bze) \mapsto (m,-1/\bze,-1/\ze) \ ,
\ee
followed by complex conjugation. The reality conditions
$\overline{\omega^+}=\omega^-$, $\overline{\omega^3}
=\omega^3$ guarantee that the two procedures are
in fact equivalent,
\be
 \overline{\tau(\Om^{[0]})} =   \Omega^{[\infty]}\ .
\ee
This is summarized by saying that the real structure $\tau$ is compatible with
the holomorphic section $\Omega$.

Finally, it can be shown that the normal bundle of the
twistor lines is isomorphic to $\IC^{2d} \times \cO(1)$.The parameter
space of the real twistor lines is therefore a manifold of real
dimension $4d$, isomorphic to the \hk manifold
$\cM$.

To summarize, any $4d$ real-dimensional \hk manifold leads to a
 $2d+1$ dimensional complex manifold $\cZ$ such that
\begin{enumerate}
\item[i)] $\cZ$ is a holomorphic fiber bundle $p:\cZ\to \IC P^1$ over the
projective line,
\item[ii)] the bundle admits a family of holomorphic sections each
  with normal bundle isomorphic to $\IC^{2d}\otimes \cO(1)$,
\item[iii)] there exists a holomorphic section $\Omega$ of
$\Lambda^2 T_F^*(2)$ defining a symplectic form on each fiber,
\item[iv)] $\cZ$ has a real structure $\tau$ compatible with i), ii),  iii)
and inducing the antipodal map on $\IC P^1$.
\end{enumerate}

Conversely, it was shown  in \cite{Hitchin:1986ea} (Theorem 3.3)
that given $\cZ,p,\tau$ satisfying i), ii), iii), iv) above, the parameter
space of real sections of $p$ is a $4d$ real-dimensional
manifold with a natural hyperk\"ahler metric for which $\cZ$ is the
twistor space. Thus, the construction of \hk metrics
reduces to the construction of $\cO(2)$-twisted holomorphic symplectic spaces.
In the rest of this section, we give a general construction of twistor spaces by patching
up local Darboux coordinate systems, generalizing the approach
in \cite{Hitchin:1986ea,Ivanov:1995cy}.

\subsection{Local sections of $\cO(m)$ on $\CP$}

We first recall the construction of the $\cO(m)$ line bundles on $\CP$.
Let $\cU_i$, $i=1\dots N$ be a set of open disks with a local coordinate
$\ze\ui{i}$. For each pair of patches, we choose an $SU(2)$ transformation
$e_{ij}$ which maps $\ze\ui{j}$ to $\ze\ui{i}$,
\be
\label{su2ij}
\ze\ui{i} = \frac{\alpha_{ij} \ze\ui{j} + \beta_{ij}}{-\bar\beta_{ij} \ze\ui{j} + \bar\alpha_{ij}}\ ,\quad
|\alpha_{ij} |^2 + | \beta_{ij} |^2 = 1 \, .
\ee
Moreover, we demand that these transformations compose properly, $e_{ij} e_{jk}=e_{ik}$.
The quantity
\be
f_{ij} (\zeta\ui{i})= \bar\beta_{ij} \ze\ui{i} + \alpha_{ij} =
\left(  -\bar\beta_{ij} \ze\ui{j} + \bar\alpha_{ij} \right)^{-1}\ ,
\ee
satisfies the cocycle condition
\be
f_{ij}(\zeta\ui{i})  f_{jk}(\zeta\ui{j}) = f_{ik}(\zeta\ui{i})\ ,\quad f_{ij}(\zeta\ui{i}) f_{ji}(\zeta\ui{j})=1\ ,
\ee
and defines the
transition function of the $\cO(1)$ line bundle on $\CP$. By definition,
holomorphic sections of the line bundle $\cO(m)$ are defined by a set of
holomorphic functions $s\ui{i}(\zeta\ui{i})$ on each patch $\cU_i$, such that, on
the overlap of two patches $\cU_i\cap \cU_j$,
\be
s\ui{i}(\zeta\ui{i})=  f_{ij} ^{m}\, s^{[j]}(\zeta\ui{j}) \, .
\ee
Equivalently, the differential form $s\ui{i}(\zeta\ui{i}) ({\rm d}\zeta\ui{i})^{-m/2}$ is
globally well defined. In particular, $\cO(-2)$ corresponds to the bundle
of holomorphic one-forms, i.e. the canonical bundle.
It is easy to see that non-zero holomorphic global sections of  $\cO(m)$ only
exist for $m\geq 0$, and that in each patch, their Taylor expansion
around $\zeta\ui{i}=0$ terminates at order $m$. In fact, the Taylor
coefficients of a global section of $\cO(m)$  form an irreducible
representation of $SU(2)$ of dimension $m+1$ (see, e.g., \cite{Ionas:2007gd}
and \cite{Alexandrov:2008nk}).

Rather than using the local coordinate systems $\zeta\ui{j}$ on each
of the patches $\cU_j$, it will be useful to single out a coordinate
$\zeta=\zeta\ui{0}$
on a particular patch $\cU_0$ referred to as the ``north pole", and extend it
to cover the whole Riemann sphere. In this coordinate, the patch $\cU_j$
is an open disk centered
at $\zeta_j=\beta_{0j}/\bar\alpha_{0j}$ (corresponding to $i=0, \ze\ui{j}=0$
in \eqref{su2ij}) and the transition function is
\be
\label{f0j}
f_{0j} = \bar\beta_{0j}  \left( \ze + 1/{\overline{\zeta_j}} \right) \ .
\ee
Note that for a given choice of $\zeta_j$, the $SU(2)$ transformations
$e_{0j}$ (and therefore the transition functions $f_{0j}$) are not uniquely
determined:
different choices correspond to different local trivializations of $\cO(1)$.
We shall abuse notation and denote by $s\ui{j}(\zeta)$
the local sections $s^{[j]}(\zeta\ui{j})$ refering to the coordinate 
$\zeta^{[j]}$ implicitly.
These local sections are defined in an open neighborhood of
$\zeta_j$, and may be extended analytically to the whole Riemann
sphere, provided no branch cuts are encountered.

Finally, for the purposes of defining the real structure, we shall
assume that the antipodal map takes each open disk $\cU_i$
with local coordinate $\ze\ui{i}$ to an open disk $\cU_{\,\bi}$  with local coordinate
$\zeta\ui{\bi}=1/{\zeta\ui{i}}$ (namely, $\alpha_{i\bi}=0, \beta_{i\bi}=\I$).
In the plane parameterized by $\zeta$, the patch
$\cU_{\bi}$ is centered at $\zeta_{\bi}=-1/\overline{\zeta_{i}}=-\alpha_{0i}/\bar\beta_{0i}$.
The patch $\cU_{\bar 0}\equiv \cU_{\infty}$
will be referred to as the ``south pole", and is related to the
north pole by the transition function $f_{0\infty}=\zeta$.
Moreover, it is easy to check that the square of the 
$\cO(1)$ transition functions satisfy the reality condition
\be
\overline{\tau(f_{ij}^2)} = f_{\bar i \bar j}^2\ .
\ee
The transition functions themselves have an inherent sign ambiguity, since the 
$\cO(1)$ bundle does not admit any real structure.

\subsection{Local Darboux coordinates and transition functions}

We now give a general construction of twistor spaces, by gluing together
patches where the holomorphic section $\Omega$ is locally trivial.
Indeed, by a trivial generalization of the Darboux theorem, it is possible to
choose local complex coordinates $(\nui{i}^I,\mui{i}_I,\ze\ui{i})$ ($I=1\dots d$)
on $\cZ$ such that the holomorphic section $\Omega$ is given locally by
\be
\label{darboux}
\Omega^{[i]}=\de\mui{i}_I\wedge \de \nui{i}^I \ .
\ee
We assume that this relation is valid in a neighborhood $\hat\cU_i$ of $\cZ$,
which projects to the open disk $\cU_i$ around the point $\ze_i$ on $\CP$.
In analogy to standard Hamiltonian mechanics, we refer to $\nui{i}^I$ and
$\mui{i}_I$ as the ``position" and ``momentum" coordinates, respectively.
Since $\Omega$ is a section of $\Lambda^2 T_F^*(2)$,
on the overlap  $\hat\cU_i \cap \hat\cU_j$, we must require
\be
\label{omij}
\Omega^{[i]}= f_{ij} ^2 \, \Omega^{[j]}  \,\quad \mod\, \de\zeta\ui{i}\ .
\ee
Thus, $(\nui{i}^I,\mui{i}_I)$ and $(\nui{j}^I,\mui{j}_I)$ must be related
by a ($\ze$-dependent) symplectomorphism. A convenient representation of such
symplectomorphisms, commonly used in Hamiltonian mechanics,
is via generating functions $\Sij{ij}$ of the initial ``position"
and final ``momentum" coordinates, whose derivatives yield the
final ``position" and initial ``momentum":\footnote{The affectation of $f_{ij}^2$ to
the momentum coordinate $\mui{i}_I$ is purely conventional at this stage.}
\be
\nui{j}^I = \p_{\mui{j}_I}\Sij{ij}(\nui{i},\mui{j},\zeta\ui{i}) \, , \qquad
\mui{i}_I =f_{ij}^2\,\p_{\nui{i}^I}\Sij{ij}(\nui{i},\mui{j},\zeta\ui{i}) \, .
\label{cantr}
\ee
To see that \eqref{cantr} preserves the twisted holomorphic symplectic
form, note that the differential of $\Sij{ij}$ is equal to the
difference of the Liouville one-forms, up to a locally exact one-form,
\be
\de\Sij{ij} = \nui{j}^I \de\mui{j}_I + f_{ij}^{-2} \mui{i}_I
\de\nui{i}^I \,\quad \mod\, \de\zeta\ui{i}\ .
\ee
Applying an exterior derivative on either sides
then proves \eqref{omij}.  In general, solving for
$(\nui{i}^I,\mui{i}_I)$ in terms of $(\nui{j}^I,\mui{j}_I)$ may lead
to ambiguities, but we assume that a prescription is given to lift
this degeneracy.

A special solution of \eqref{omij} is to assume that $\nui{i}^I$ and $\mui{i}_I$
separately transform as sections of $\cO(2n)$ and $\cO(2-2n)$, such that their
product transforms as a section of $\cO(2)$:
\be
\label{o2o2}
\nui{j}^I = f_{ij}^{-2n}\nui{i}^I\ ,\quad
\mui{j}_I = f_{ij}^{2n-2}\mui{i}_I\ ,\quad
\ee
corresponding to the generating function
\be
\label{o2o2s}
\Sij{ij}(\nui{i},\mui{j},\zeta\ui{i})=f_{ij}^{-2n}\nui{i}^I\mui{j}_I \ .
\ee
We shall assume that the generating function \eqref{o2o2s} controls the
``microscopic" structure of the twistor space: in other words, given an
open covering by patches $\cU_i$ with non-trivial symplectomorphisms
$\Sij{ij}$, we assume that any refinement of the covering to include
a new patch  $\cU_j\subset \cU_i$ is controlled by the trivial gluing function
\eqref{o2o2s}.

Of course, the transition functions $\Sij{ij}$ are subject to
consistency conditions: in particular $\Sij{ji}$ should be the
generating function of the inverse of the symplectomorphism generated
by $\Sij{ij}$, determined by the Legendre transform of $\Sij{ij}$:
\be
\Sij{ji}(\nui{j},\mui{i},\zeta\ui{j})=
\langl \nui{i}^I\mui{i}_I+f_{ij}^2\,\nui{j}^I\mui{j}_I
-f_{ij}^2 \,\Sij{ij}(\nui{i},\mui{j},\zeta\ui{i})\rangl_{\nui{i},\mui{j}}\, .
\label{consist_inv}
\ee
Here, $\langle\cdot\rangle_{x}$ denotes the result of extremising with
respect to the variable $x$. When several extrema exist,
we assume again that an additional prescription is provided.
Moreover, the composition of the symplectomorphisms generated by  $\Sij{ik}$
and  $\Sij{kj}$ should be the  symplectomorphism generated by  $\Sij{ij}$,
\be
\Sij{ij}(\nui{i},\mui{j},\zeta\ui{i})=
\langl f_{jk}^2\, \Sij{ik}(\nui{i},\mui{k},\zeta\ui{i})
+\Sij{kj}(\nui{k},\mui{j},\zeta\ui{k})
-f_{jk}^2\, \nui{k}^I\mui{k}_I\rangl_{\nui{k},\mui{k}}\, .
\label{consist_cond}
\ee
Finally, the Darboux form \eqref{darboux} of the section $\Om$ on the patch $\cU_i$
does not fix the coordinates $(\nui{i},\mui{i})$ uniquely: it is still possible to
perform a local symplectomorphism  $(\nui{i},\mui{i})\to (\tnui{i},\tmui{i})$
generated by a holomorphic function $\Ti{i}(\nui{i},\tmui{i},\zeta\ui{i})$,
regular in the patch $\hat\cU_i$,
\be
\tnui{i}^I = \p_{\tmui{i}_I}\Ti{i}(\nui{i},\tmui{i},\zeta\ui{i}) \, , \qquad
\mui{i}_I =\p_{\nui{i}^I}\Ti{i}(\nui{i},\tmui{i},\zeta\ui{i}) \, .
\label{cantrloc}
\ee
Therefore, the set of transition functions $\Sij{ij}$ defines the same
holomorphic symplectic space as
\bea
\tSij{ij}(\nui{i},\mui{j},\zeta\ui{i})&=&
\langl f_{ij}^{-2}\(\Ti{i}(\nui{i},\tmui{i},\zeta\ui{i})-\tnui{i}^I\tmui{i}_I\)
+ \Sij{ij}(\tnui{i},\tmui{j},\zeta\ui{i})
\right.
\nonumber \\
&& \qquad\qquad \left.
-\tnui{j}^I\tmui{j}_I+\Tiinv{j}(\tnui{j},\mui{j},\zeta\ui{j})\rangl_{\tnui{i},\tmui{i},\tnui{j},\tmui{j}}\, ,
\label{gaugefr}
\eea
where $\Tiinv{i}$ is the generating function of the inverse of the symplectomorphism
\eqref{cantrloc}, given by a formula similar to \eqref{consist_inv}:
\be
\Tiinv{i}(\tnui{i},\mui{i},\zeta\ui{i})=
\langl \nui{i}^I\mui{i}_I+\tnui{i}^I\tmui{i}_I
-\Ti{i}(\nui{i},\tmui{i},\zeta\ui{i})\rangl_{\nui{i},\tmui{i}}\, .
\ee

To summarize, a holomorphic symplectic manifold $\cZ$ covered by $N$
patches $\hat\cU_i$ is uniquely specified by a set of $N-1$ holomorphic
functions $\Sij{0i}$ of $2d+1$ variables $\nu^I, \mu_I, \zeta$, subject
to the equivalence relation $\Sij{ij}\simeq \tSij{ij}$ in \eqref{gaugefr}.
The rest of the transition functions is determined by the composition
and inversion rules \eqref{consist_cond} and \eqref{consist_inv}. With this
data in hand, one may in principle solve \eqref{cantr} for all pairs of overlapping patches,
under the assumption that  $\nui{i}^I$ and $\mui{i}_I$ are regular at $\zeta\ui{i}=0$,
and determine all $\nui{i}^I$ and $\mui{i}_I$ in each patch as functions of
$\zeta$ and an a priori unspecified number $d'$ of complex parameters.

While it is useful to discuss the complex and real structures separately,
it should be noted at this point that the gauge freedom \eqref{gaugefr}
may be used to bring the real structure in a standard form,
\be
\overline{\tau\bigl(\nui{i}^I\bigr)}=-\nui{\bi}^I\, ,
\qquad
\overline{\tau\bigl(\mui{i}_I\bigr)}=-\mui{\bi}_I \, .
\label{realcon}
\ee
Here we have assumed that $n$ is integer;
cases with half-integer $n$ can be treated similarly, by imposing
(in physics parlance,  symplectic Majorana) reality conditions on pairs
of $\cO(2n)$ multiplets. The reality conditions \eqref{realcon}
require that the transition functions $\Sij{ij}$ satisfy
\be\label{Sreal}
\overline{ \tau\(\Sij{ij}(\nui{i}, \mui{j}, \zeta\ui{i}) \)} =
\Sij{\bi \bj}(\nui{\bi}, \mui{\bj}, \zeta\ui{\bi}) \, .
\ee
In the particular case $j=\bi$,
taking into account that $\Sij{\bi i}$ is related to $\Sij{i\bi}$ by \eqref{consist_inv},
we see that the reality condition connects the antipodal map of this transition function
with its Legendre transform.
Taking into account these reality conditions, the $d'$
complex parameters of the twistor lines mentioned above thus become $d'$ real
parameters.
The assumption ii) of the Theorem implies that $d'=4d$, and that the
parameter space is in fact the base $\cM$ itself.
The functions $\nui{i}^I(\zeta)$ and $\mui{i}_I(\zeta)$  provide a parametrization
of the ``twistor lines", i.e. of the fiber over
a point $m$ in $\cM$ as a rational curve in $\cZ$.

Having determined the Darboux coordinates $(\nui{i}^I,\mui{i}_I)$ as
functions of $\ze$ and of the $4d$ parameters, it is now
straightforward to compute the metric: complex coordinates on $\cM$ in
the complex structure $J(\ze_i,\bze_i)$ are given by evaluating the
twistor lines $(\nui{i}^I,\mui{i}_I)$ at $\ze\ui{i}=0$ (or
equivalently $\zeta=\ze_i$),
\be \vi{i}^I =\nui{i}^I(\zeta_i)\ ,\quad
\wi{i}_I=\mui{i}_I (\zeta_i)\ ,
\ee while the \kahler form
$\om(\ze_i,\bze_i) \equiv\om^{3[i]}$ is given by Taylor expanding the holomorphic
section $\Om^{[i]}$ around $\ze\ui{i}=0$:
\be
\Om^{[i]} (\zeta\ui{i}) =  \de\wi{i}_I \wedge \de\vi{i}^I -\I \om^{3[i]} \ze\ui{i}
+\de\bwi{i}_I \wedge \de\bvi{i}^I  (\zeta\ui{i})^2\ .
\ee
Knowing the complex coordinates and the \kahler form, it is then
straightforward to obtain the metric. A \kahler potential
$K^{[i]}(\vi{i},\bvi{i},\wi{i},\bwi{i})$ may be computed by
integrating the \hk form $\om^{3[i]}$.  It will in general depend on
the patch $i$; a notable exception is the case of \hk cones,
discussed in  \cite{Alexandrov:2008nk}, where it is possible to find a single ``\hk
potential'' $\chi$ which can serve as a \kahler potential for all
complex structures at once.

The prescription given just above only applies to the complex
structures corresponding to the marked point $\zeta_i$ in the patch
$\cU_i$. The complex structure at a point $\zeta_j$ lying in the patch $\cU_i$, but
distinct from all $\zeta_k$, $k=1\dots N$ may be described by refining the
open covering and adding a new patch $\cU_j$ centered at $\zeta_j$, as
indicated below \eqref{o2o2s}.

\subsection{A remark on holomorphic quantization}

 As we have indicated repeatedly, the construction of the twistor space $\cZ$
 via a set of symplectomorphisms relating local Darboux coordinate systems
 has strong similarities to classical Hamiltonian mechanics, albeit in
 a holomorphic setting.  It is tempting to ask whether this classical construction
 can be quantized. The following construction naturally suggests itself:
 to any patch $\cU_i$, associate a Hilbert space $\mathcal{H}^{[i]}$ of holomorphic functions
 $\Psi^{[i]}(\nui{i}^I,\ze\ui{i})$. On the overlap $\hat\cU_i \cap \hat\cU_j$, require that
 \be
 \label{intert}
 \Psi^{[i]}(\nui{i},\ze\ui{i})
 \sim  f_{ij}^m\, \int\, \de^d \mui{j}_I \, \de^d \nui{j}^I \, \exp\left[ \frac{\I}{\hbar}
 \left( \Sij{ij}( \nui{i},\mui{j},\ze\ui{i} ) -
 \mui{j}_I \nui{j}^I \right) \right]\,
 \Psi^{[j]}(\nui{j},\ze\ui{j})\ ,
 \ee
 where $m$ is an undetermined parameter, and $\hbar$ is Planck's constant.
 Note that for a trivial transition function \eqref{o2o2s}, the integral
 over $\mui{j}_I$ imposes the first equation in \eqref{o2o2} as a delta
 function, leading to
 \be
 \label{intertt}
 \Psi^{[i]}(\nui{i},\ze\ui{i})
 \sim f_{ij}^m\,
 \Psi^{[j]}( f_{ij}^{-m} \nui{i},\ze\ui{j}) \ .
 \ee
 Thus $\Psi$ is valued in $\cO(m)$ within each patch, but may transform
 non-trivially from one patch to another.
 The right hand side of \eqref{intert}
 is the standard semi-classical form of the intertwiner between wave functions
 in polarizations related by a symplectomorphism generated by $S$, and
 is only expected to be correct in the limit $\hbar\to 0$. For finite values
 of $\hbar$,
 it should be corrected in order for
 the intertwiners from $\cU_i$ to $\cU_j$ and from $\cU_j$ to $\cU_k$ to
 compose properly. Assuming that this quantum ambiguity can be fixed,
  \eqref{intert} defines a global Hilbert space $\mathcal{H}_k$ which can be
  viewed as the holomorphic quantization of the space $\cZ$. When $\cM$
  is the Swann bundle \cite{MR1096180,deWit:2001dj} of the hypermultiplet moduli space, it is tempting
  to think that $\mathcal{H}_m$, for a suitable value of $m$, is the habitat
  of the generalized topological amplitude postulated in \cite{Gunaydin:2006bz}.

\section{Twistor spaces of $\cO(2n)$ \hk manifolds}
\label{sec_caseisom}

We now restrict to the case of HK manifolds obtainable by the generalized Legendre
transform method \cite{Hitchin:1986ea,Lindstrom:1987ks}. As shown in \cite{bielawski2000tqh},
these manifolds are characterized by the fact that they admit a local Hamiltonian action of
a $d$-dimensional abelian twistor group (i.e., a Lie group whose parameters are sections of
$\cO(2-2n)$), or equivalently, by the existence of a covariantly-constant, tri-holomorphic
higher-rank tensor. For $n=1$, this reduces to the standard Legendre transform
construction  \cite{Hitchin:1986ea} of $4d$-dimensional HK manifolds with $d$
commuting tri-holomorphic isometries. The twistorial interpretation of these
constructions was explained in \cite{Hitchin:1986ea,Ivanov:1995cy}, in a 
set-up where the twistor space is covered by two patches only, projecting to the north
and south pole of $\CP$. In this section, we extend their analysis to the case of
an arbitrary number of patches $\cU_i$. This extension is needed in order to 
determine the solutions for the twistor lines (in particular 
the $\mu_I(\zeta)$) defined on the entire twistor space $\cZ$.

\subsection{Symplectomorphisms compatible with the group action}

As explained in \cite{bielawski2000tqh}, the existence of a
Hamiltonian action of a $d$-dimensional abelian twistor group implies that
there exist $d$ global sections of $\cO(2n)$, corresponding to the moment
map\footnote{Recall that if $V$ is a Hamiltonian vector field, $\cL_V \Omega
=(d \imath_V+\imath_V d) \Omega=0$
so $\imath_V \Omega=d\mu_V$, where $\mu_V$ is the moment map of $V$,
defined up to an additive constant.
If $V$ is twisted by $\cO(2-2n)$ and $\Omega$ by $\cO(2)$, then
$\mu_V$ is twisted by $\cO(2n)$.}   of these actions.
Since these actions commute, we may use them
as ``position'' coordinates $\nu^I$. The fact that they are globally well-defined
means that, on  $\hat\cU_i \cap \hat\cU_j$,
\be
\nui{j}^I = f_{ij}^{-2n}\nui{i}^I \ .
\label{oncantr1}
\ee
In order for \eqref{omij} to hold, the conjugate coordinates $\mu_I$ must
transform as
\be
\mui{i}_I =f_{ij}^{2-2n}\mui{j}_I
-f_{ij}^2\p_{\nui{i}^I}\tHij{ij}(\nui{i},\zeta\ui{i}) \ ,
\label{oncantr2}
\ee
where $\tHij{ij}$ may be a function of $\zeta\ui{i}$ and the positions $\nui{i}^I$,
but cannot depend on the momenta $\mui{j}_I$. This symplectomorphism
is of the form \eqref{cantr} for a special choice of generating function
\be
\Sij{ij}(\nui{i},\mui{j},\zeta\ui{i})=f_{ij}^{-2n}\nui{i}^I\mui{j}_I
-\tHij{ij}(\nui{i},\zeta\ui{i}) \, .
\label{trans_2n}
\ee
Note that the case $\tHij{ij}=0$ reduces to the transformation rules \eqref{o2o2}.
The compatibility constraints \eqref{consist_inv} and \eqref{consist_cond} require that
\be
\tHij{ji}(\nui{j},\zeta\ui{j}) = -f_{ij}^{2} \tHij{ij}(\nui{i},\zeta\ui{i})
\ ,
\ee
and
\be
\tHij{ik}(\nui{i},\zeta\ui{i}) + f_{kj}^2 \tHij{kj}(\nui{k},\zeta\ui{k})=
 f_{kj}^2 \tHij{ij}(\nui{i},\zeta\ui{i}) \, .
\ee

As explained above \eqref{f0j}, it is convenient to single out the patch $i=0$,
and use the complex coordinate $\zeta\equiv \zeta\ui{0}$ as a global
coordinate on $\cZ$, allowing for poles in the $\zeta$ plane (branch
cuts will be discussed in Section \ref{subsec_branch}).
Moreover, since $\nu$ is globally well-defined, we may trade the
argument $\nui{i}$ in $\tHij{ij}$ for
\be
\eta^I(\zeta) \equiv \zeta^{-n}\nui{0}^I(\zeta) = \sum\limits_{k=-n}^{n} \eta^I_{k}\,\zeta^k \, ,
\label{defeta}
\ee
and  define
\be
\label{HHt}
\Hij{ij}(\eta,\zeta)\equiv \zeta^{-1}f_{0j}^2
\,\tHij{ij}(\zeta^n f_{0i}^{-2n}\eta,\zeta)\, .
\ee
In \eqref{defeta} and \eqref{HHt}, the factors of $\zeta^{-n}$ are conventional,
and are inserted to facilitate comparison with earlier studies.
In terms of $\Hij{ij}$, the relation \eqref{oncantr2}
becomes
\be
\mui{i}_I=f_{ij}^{2-2n}\mui{j}_I-\zeta^{1-n}f_{0i}^{2n-2} \,\p_{\eta^I}\Hij{ij}(\eta,\zeta)\, .
\label{eq2n}
\ee
Thanks to the redefinition \eqref{HHt}, the consistency conditions
\eqref{consist_cond} take the simple form
\be
\Hij{ji}=-\Hij{ij}, \qquad \Hij{ik}+\Hij{kj}=\Hij{ij}\, ,
\label{consist_condH}
\ee
where all functions depend on the same variables $\eta^I$ and $\zeta$.
Similarly, the gauge freedom \eqref{gaugefr} with
\be
T\ui{i}(\nui{i},\tmui{i},\zeta\ui{i})= \nui{i}^I \tmui{i}_I - \zeta f_{0i}^{-2} \, G\ui{i}(\eta,\zeta)\ ,
\ee
allows to shift
\be
\Hij{ij}\mapsto \Hij{ij}+\Gi{i}-\Gi{j}\, ,\quad 
\mui{i}_I \mapsto \mui{i}_I  + \zeta^{1-n}f_{0i}^{2n-2} \,\p_{\eta^I}\Gi{i}\ .
\label{gaugeHG}
\ee
It is important that $\zeta^{1-n}f_{0i}^{2n-2} \,\p_{\eta^I}\Gi{i}$ be a regular 
function of $\zeta$ in the patch $\hat\cU_i$.
Thus, the set of $\Hij{ij}$ should be viewed as a class in the Cech cohomology group $H^1(\cZ)$
valued in the set of $\mu$-independent holomorphic functions.
We shall often abuse notation and define $\Hij{ij}$ away from the overlap $\hat\cU_i \cap \hat\cU_j$
(in particular when the two patches do not intersect) using analytic continuation and the 
second equation in \eqref{consist_condH} 
to interpolate from $\hat\cU_i$ to $\hat\cU_j$. Ambiguities in the
choice of path can be dealt with on a case by case basis.

As in \eqref{Sreal}, reality conditions restrict the possible functions $\Hij{ij}$.
The conditions \eqref{realcon} on $\nu^I$ translate into
\be
\tau(\eta^I)=(-1)^{n-1}\bar {\eta}^I  \quad \mbox{i.e.}\quad \bar\eta^I_{-k} =(-1)^{n+k-1}\eta^I_k \ .
\label{realeta}
\ee
Then, the reality conditions on $\mu_I$ require that
\be
\overline{\tau(\Hij{ij})} = -\Hij{\bi\bj}\, .
\label{realH}
\ee
In particular, the transition function relating a patch with its antipodal map must be invariant,
$\overline{\tau(\Hij{i\bi})} = \Hij{i\bi}$. These conditions ensure that the metric and \kahler
potential obtained in the following subsections are real.

\subsection{\kahler potential and twistor lines for $\cO(2n)$ manifolds, $n\geq 2$}
\label{seco2n}

We now proceed to find the twistor lines, i.e. find the most general functions
$\eta^I(\zeta)$ and $\mui{i}_I(\zeta)$ satisfying \eqref{eq2n}, subject to the
requirement that the expansion of $\eta^I$ in \eqref{defeta} has only $2n+1$ terms,
while each $\mui{i}_I$ is regular at $\zeta=\zeta_i$.  According
to the Theorem in \ref{sec-rev} and for suitably generic choices of functions $\Hij{ij}$ obeying hypothesis ii), the parameter space of these twistor lines will
be a $4d$ dimensional \hk manifold with an action of a $d$-dimensional twistor
group.
Clearly, the case $n=1$ is quite different from
$n\geq 2$: in the latter case, the number of coefficients $\eta^I_k$
in \eqref{defeta} is greater than $4d$ and solutions to  \eqref{eq2n}
are expected to exist only when $(2n-3)d$ constraints are imposed on $\eta^I_k$.
In contrast, for $n=1$ the $3d$ coefficients $\eta^I_k$ are expected to be
unconstrained, and supplemented by $d$
additional parameters coming from the sections $\mui{i}_I$:
indeed, for $n=1$ Eq. \eqref{eq2n} is manifestly invariant under
global shifts  $\mui{i}_I(\zeta) \to \mui{i}_I(\zeta) + \vrh_I$, corresponding to
the $d$ tri-holomorphic isometries.
In this subsection, we shall restrict our attention to the case $n\geq 2$,
postponing the case $n=1$ to Subsection 3.3.

Let us consider the Taylor expansion of $\mui{0}_I(\zeta)$ around the north pole.
The $k$-th coefficient can be obtained from the contour integral
\be
\mu\ui{0}_{I,k} = \frac{1}{2\pi\I} \oint_{C_0} \frac{\de\zeta}{\zeta^{1+k}} \mu\ui{0}_I(\zeta)\ ,
\label{muinfk}
\ee
where $C_0$ is a contour around 0 inside $\cU_0$, oriented counterclockwise.
The contour $C_0$ may be deformed into a sum of contours inside each $\cU_j$.
Using \eqref{eq2n} with $i=0$ in each patch, we obtain
\be
\mu\ui{0}_{I,k} = - \frac{1}{2\pi\I} \sum_{j\neq 0}
\oint_{C_j} \frac{\de\zeta}{\zeta^{1+k}}
\left[ f_{0j}^{2-2n}\mui{j}_I-\zeta^{1-n}\,\p_{\eta^I}\Hij{0j}(\eta,\zeta) \right]\ .
\label{muinfk2}
\ee
Recall from \eqref{f0j} that $f_{0j}$ has a zero at $\zeta= -1/{\overline{\zeta_j}}$
and a pole at $\zeta=\infty$. Therefore, for $j\neq \infty$, the first term inside the bracket is
regular inside the contour $C_j$ and can be omitted. For $j=\infty$, since $f_{0\infty}=\zeta$,
the first term picks the $(2-2n-k)$-th coefficient in the Taylor expansion of
$\mui{\infty}_I$ around infinity. Thus, we get
\be
\label{dereq}
\mu\ui{0}_{I,k} =  \mui{\infty}_{I,2n-2-k} +
\frac{1}{2\pi\I} \sum_{j}\oint_{C_j} \frac{\de\zeta}{\zeta^{k+n}}
\,\p_{\eta^I}\Hij{0j}(\eta,\zeta) \, ,
\ee
where the sum over $j$ was trivially extended to include 0. This equation
holds for all values of $k$. However, $\mu\ui{0}_{I,k}$ vanishes if $k<0$,
and $\mui{\infty}_{I,2n-2-k}$ vanishes if $k>2-2n$. We conclude that,
for $|\ell |\le n-2$,
\be
\sum_{j}\oint_{C_j} \frac{\de\zeta}{2\pi \I\,\zeta^{1+\ell}}
\pa_{\eta^I} \Hij{0j}(\eta,\zeta)  = 0,
\label{conwzero}
\ee
while, for $k\geq 0$,
\be
\label{contintrep}
\muor_{I,k} =
  \sum_{j}\oint_{C_j}  \frac{\de\zeta}{2\pi \I\,\zeta^{n+k}}
\pa_{\eta^I} \Hij{0j}(\eta,\zeta) \ ,\quad
\muinf_{I,k} =
-  \sum_{j}\oint_{C_j}  \frac{\de\zeta}{2\pi \I\,\zeta^{2-n-k}}
\pa_{\eta^I} \Hij{0j}(\eta,\zeta)  \ .
\ee
The sections $\muor_{I}$ and $\muinf_{I}$ may be obtained
in terms of the coefficients of $\eta$ by resumming the series,
\be
\muor_{I}(\zeta) =
  \sum_{j}\oint_{C_j}  \frac{\de\zeta'}{2\pi \I(\zeta')^{n}} \frac{\zeta'}{\zeta'-\zeta}
\pa_{\eta^I} \Hij{0j}(\eta,\zeta)\ ,
\ee
\be
\muinf_{I}(\zeta) =
  \sum_{j}\oint_{C_j}  \frac{\de\zeta'}{2\pi \I\,(\zeta')^{2-n}} \frac{\zeta}{\zeta'-\zeta}
\pa_{\eta^I} \Hij{0j}(\eta,\zeta)\ ,
\ee
where the first (resp. second) equation holds at the north (resp. south) pole.

On the other hand, the conditions \eqref{conwzero} give $(2n-3)d$ constraints
on the $(2n+1)d$ coefficients of the $\eta^I$, leaving generally a $4d$-dimensional
set of solutions. These constraints are best described by introducing
the ``Lagrangian''\cite{Karlhede:1984vr,Lindstrom:1987ks}
\be
\label{tlag}
\mathcal{L} (\eta_k)=\sum_j
\oint_{C_j} \frac{\de\zeta}{2\pi \I\,\zeta}\, \Hij{0j}(\eta,\zeta)\, .
\ee

Note that due to the consistency conditions \eqref{consist_condH}, the index
$0$ on the right-hand side of this expression may be substituted with any
other value without changing the result (however,  \eqref{tlag} is
adapted to the complex structure at $\zeta=0$, and to that complex structure
only). In terms of $\CL$, the constraints \eqref{conwzero} become
\be
\pa_{\eta^I_{-k}}
\mathcal{L}=0, \qquad |k|\le n-2\ ,
\label{conwzeroL}
\ee
and in principle allow to eliminate the coefficients $\eta^I_{-k}$, $ |k|\le n-2$.
The remaining coefficients
\be
v^I\equiv \eta_{-n}^I\, ,
\quad
x^I\equiv \eta^I_{1-n}\, ,
\quad
\bv^I\equiv -\eta_{n}^I\, ,
\quad
\bx^I\equiv \eta^I_{n-1}\ ,
\label{def_variab}
\ee
parametrize the twistor lines. The compatibility with the real structure requires
that $\bv^I$ and $\bx^I$ be complex conjugate to $v^I, x^I$.

Expanding the holomorphic section $\Omega$
around $\zeta=0$, one easily finds that the holomorphic
symplectic form is given by
 \be
 \label{om1vw}
\omega^+ = \de w_I \wedge \de v^I \, ,
\ee
where
\be
\qquad
w_I\equiv \muor_{I,0}\, ,\qquad \bw_I\equiv -\muinf_{I,0}\, .
\ee
Thus, $(v^I, w_I)$ form a system of holomorphic Darboux
coordinates for the complex structure $J^3$. The parameters
$x^I, \bar x^I$ may be eliminated in favor of $(v^I, w_I)$
using Eq. \eqref{contintrep} for $k=0$, which amounts to
\be
\label{dxLw}
\pa_{x^I} \mathcal{L} = w_I\, .
\ee
To next order in $\zeta$, using
\be\label{muzeta1}
\begin{split}
\muor_I(\zeta) = & \, w_I + \zeta\,\pa_{v^I}\mathcal{L}\,  + \mathcal{O}(\zeta^2)\, , \\
\muinf_I(\zeta) = & \, -\bar w_I +  \zeta^{-1} \,\pa_{\bv^I}\mathcal{L}
+ \mathcal{O}(\zeta^{-2}) \, ,
\end{split}
\ee
one finds that the K\"ahler form $\omega^3$ descends
from a K\"ahler potential
\be
K(v^I,\bv^I,w_I,\bw_I)
= \langle \mathcal{L}\(v^I,\bv^I,x^I,\bx^I,\eta_k^I\) - x^I w_I- \bx^I \bw_I \rangle_{x^I,{\bar x}^I,\eta^I_k}  \, .
\label{Legtr}
\ee
The conditions \eqref{conwzeroL} and \eqref{dxLw} imply that the
right-hand side of \eqref{Legtr} is extremized with respect to $x^I,\bx^I,\eta_k^I$, as denoted by the brackets.
Thus, the K\"ahler potential is the Legendre transform of the Lagrangian \eqref{tlag}
(at vanishing momentum conjugate
to $\eta_k$). The reality conditions \eqref{realH} guarantee that $K$ is real.
The complex coordinates and K\"ahler potential
for other choices of complex structure may be obtained by following
the same chain of reasoning, but singling out a different ``north pole''.

\subsection{\kahler potential and twistor lines for $\cO(2)$ manifolds}

We now turn to the case $n=1$, corresponding to toric HK manifolds
with $d$ commuting triholomorphic isometries. As indicated previously, the
$\cO(2)$ global sections $\eta^I$ are unconstrained; we denote their
Laurent coefficients at $\zeta=0$ as
\be
\eta^I(\zeta) = \frac{v^I}{\zeta} + x^I - \bv^I \zeta\, .
\label{omult}
\ee
This is the $N=2$ tensor multiplet in projective superspace, as introduced
in \cite{Karlhede:1984vr}. On the other hand, local sections $\mui{i}_I$ should satisfy \eqref{eq2n},
\be
\mui{i}_I=\mui{j}_I-\p_{\eta^I}\Hij{ij}(\eta,\zeta) \ .
\label{eq22}
\ee
These equations are  expected to uniquely determine all the $\mui{i}_I$'s
(at least locally), up to global shifts $\mui{i}_I\to \mui{i}_I +\vrh_I$.
Indeed, \eqref{eq22} can be solved following the same steps as
in the previous subsection.  Setting $n=1$ in  \eqref{dereq},
Eq. \eqref{contintrep} continues
to hold for $k>0$.  For $k=0$ however, only the difference
$\muor_{I,0}-\muinf_{I,0}$ is determined,
\be
\muor_{I,0}-\muinf_{I,0} =\sum_j \oint_{C_j} \frac{\de\zeta}{2\pi \I\,\zeta}\,
\pa_{\eta^I} H^{[0j]}(\eta,\zeta)\, .
\label{muzeroisom}
\ee
Together with $v^I, x^I, \bv^I$, the real combinations
\be
\vrh_I\equiv -\I ( \muor_{I,0}+\muinf_{I,0} ) \ ,
\ee
provide the $4d$ coordinates parametrizing the solution space of \eqref{eq22}.
Defining as before $w_I\equiv \muor_{I,0}, \bw_I\equiv -\muinf_{I,0}$,
we see that $w_I+\bw_I$ is the parameter conjugate to $x^I$ with
respect to the Lagrangian \eqref{tlag},
\be
\label{tlag1}
w^I + \bw^I =\pa_{x^I} \cL \, ,\qquad
\mathcal{L} (v^I, \bv^I,x^I)=\sum_j
\oint_{C_j} \frac{\de\zeta}{2\pi \I\,\zeta}\, \Hij{0j}(\eta^I,\zeta)\, .
\ee
As before, $(v^I, w_I)$ provide a system of holomorphic Darboux coordinates
on $\cM$ in the complex structure $J^3$. The holomorphic symplectic form
is still given by \eqref{om1vw}, while the Legendre transform of the Lagrangian
$\cL$ produces a \kahler potential for a \hk metric in this complex structure,
\be
K(v, \bv,w, \bw) = \langle \cL(v^I,\bv^I,x^I) - x^I  (w_I + \bw_I) \rangle_{x^I} \ .
\label{K02}
\ee
The independence of the \kahler potential on $\I \vrh_I=w^I-\bw^I$ makes it
manifest that the metric has $d$ commuting Killing vectors $\pa_{\vrh_I}$,
which furthermore are tri-holomorphic. The metric on $\cM$ may be computed
without knowing $x^I$ as an explicit function of $(v^I,\bv_I, w_I + \bw_I)$. 
It is given by \cite{Hitchin:1986ea}
\be\label{HKmet0}
\begin{split}
K_{v^I \bv^J} =& \, \CL_{v^I \bv^J}-\CL_{v^I x^K}\CL^{x^K x^L}\CL_{x^L \bv^J} \, ,
\\
K_{v^I \bw_J} =& \, \CL_{v^I x^K}\CL^{x^K x^J}\, , \quad
K_{w_I \bv^J} =\, \CL^{x^I x^K}\CL_{x^K \bu^J} \, ,\quad
K_{w_I \bw_J} =-\CL^{x^I x^J}\ ,
\end{split}
\ee
where  $\CL_{x^I}=\p_{x^I}\CL$, $\CL_{v^I}=\p_{v^I}\CL$, and
$\CL^{x^Ix^J}$ denotes the inverse of the matrix $\CL_{x^Ix^J}$.

The local sections $\muor_I$ can be computed by resumming the series
\eqref{contintrep}, leading to
\be
\muor_I =  w_I + \frac{\pa}{\pa v^I}  \left[
\sum_j \oint_{C_j} \frac{\de\zeta'}{2\pi \I}
\,\frac{\zeta \, \Hij{0j}(\zeta')}{\zeta'-\zeta}
  \right]\ ,
\label{solmu}
\ee
when $\zeta$ lies inside the north pole patch $\cU_0$, or
\be
\muinf_I =  -\bw_I - \frac{\pa}{\pa \bv^I}  \left[
\sum_j \oint_{C_j} \frac{\de\zeta'}{2\pi \I\zeta'}
\,\frac{\Hij{0j}(\zeta')}{\zeta'-\zeta}
  \right]\ ,
\label{solmub}
\ee
when $\zeta$ lies in the south pole patch $\cU_\infty$. Inserting $w_I=(\cL_{x^I} +\I  \vrh_I)/2$ in
\eqref{solmu}, or $\bw_I=(\cL_{x^I} -\I  \vrh_I)/2$ in \eqref{solmub}, leads
to a single expression
\be
\mui{i}_I (\zeta)=
\frac{\I}{2}\, \vrh_I +\sum_{j} \oint_{C_j} \frac{\de\zeta'}{2\pi \I\,\zeta'}\,
\frac{\zeta+\zeta'}{2 (\zeta'-\zeta)}\,
 \p_{\eta^I}H^{[0j]}(\zeta'),
\label{mui}
\ee
valid both for $i=0,\zeta\in \cU_0$ and $i=\infty, \zeta\in \cU_\infty$.
In fact, it is easy to check that \eqref{mui} evaluated for $\zeta\in \cU_i$
gives the general solution to the matching conditions \eqref{eq22}.
Equation \eqref{mui} is the main new result of this section.
It should be noted that the index 0 on the right hand side may be substituted with
any other value without changing the result, due to the consistency conditions
\eqref{consist_condH}. Eqs.  \eqref{omult}  and \eqref{mui}
provide an explicit parametrization of the twistor lines over any point on
$\cM$ as rational curves in $\cZ$.

\subsection{Logarithmic branch cuts}
\label{subsec_branch}

Up till now we assumed that the local coordinates $\nui{i}(\ze)$ and $\mui{i}(\ze)$ could
be analytically extended as meromorphic functions on $\CP$. In many cases of
interest however, it is important to relax this assumption and allow for branch cuts.
In this subsection we extend the previous construction to the case where the
transition function $\Hij{ij}$ contains a singular term $\cij{ij} \log\eta$,
where  $\eta$ is one of the $\cO(2)$ multiplets $\eta^I$ and $\cij{ij}$ are regular functions of these multiplets. Correspondingly, one
expects a singularity in the associated local momentum coordinates $\mui{i}$. An example of
this situation is the ``improved tensor multiplet" representation of flat $\IR^4$,
discussed in Section \ref{subsec_secondflat} below, and the one-loop corrected
hypermultiplet moduli space, to be discussed in \cite{Alexandrov:2008nk}.

The presence of the logarithmic branch cut gives rise to a Riemann surface which is
an infinite cover of $\CP$. Let us label each sheet by an integer, and single out one
particular sheet, say with the label $n$. As before,  this sheet can be covered by a
set of patches $\cU_i$, with local Darboux coordinates $\nui{i}^I,\mui{i}_I$
satisfying  \eqref{eq22} on the overlap of two patches. In addition, we must make a
choice of branch for $\log\eta$. At this point, it is important to recall that logarithmic
branch cuts are oriented: depending whether the angle between the branch cut and the path is positive or negative, one passes from the $n$-th sheet to the
$n+1$-th or $n-1$-th sheet. We choose the branch cuts for the singular
term in $\Hij{ij}$ to extend from $\ztp$ to the origin
and from $\ztm$ to infinity, where $\zeta_\pm$ are the two zeros of
$\eta$ (of course one can flip $\ztp$ and $\ztm$).

In this situation, the procedure to solve the equations for $\mui{i}$ presented in
Section \ref{seco2n} continues to hold, with due attention paid to the cuts.
Since $\muor$ is regular at $\zeta=0$, its Taylor coefficients are still given
by \eqref{muinfk}. The integral over $C_{0}$ can again be rewritten as a
sum of integrals around contours surrounding all $\cU_i$. Now,
however, contours can encounter cuts of $\muor$
and therefore may not be individually closed.
For \eqref{muinfk2} to hold, we require that the end point of one contour
lies on the branch cut, and coincides with the starting point of another contour, in
such a way that the sum reproduces a closed contour (see Fig. \ref{figure0}).
Using \eqref{muinfk2}, one may substitute $\muor$ in the patch $\cU_j$ by
$\mui{j}-\pa_{\eta}\Hij{0j}$ as in  \eqref{eq22}; the first term is regular
inside $\cU_j$, so reproduces the first term in \eqref{dereq}. The second
term can be rewritten as the integral of a holomorphic function on a
closed contour, due to additional conditions on the transition functions.

Indeed, let us analyze the possible structure of singularities in $\mui{i}$ and its implications
for the coefficients $\cij{ij}$. The first possibility is that $\mui{0}$ does not have
logarithmic singularities. Then they can appear
only in the transition functions $\Hij{0j}$ for $j\ne \pm,\infty$. In this case however, the
corresponding contours $C_j$ do not cross the branch cuts and therefore can be closed.
The second possibility is that $\mui{0}$ does contain branch cut singularities which,
under our assumptions, must lie at $\zeta=\ztpm$ and infinity.
Since singularities must appear in pairs, the discussion can be reduced to
only two cases: (i) $\mui{0}$ has a singular term of the form $\log(\zeta-\ztm)$,
corresponding to a branch cut from $\ztm$ to $\infty$,
or (ii) $\mui{0}$ has a singular term of the form $\log\frac{\zeta-\ztm}{\zeta-\ztp}$,
corresponding to a branch cut from $\ztm$ to  $\ztp$. Since the addition of the
transition function $\p_{\eta}\Hij{0j}$ must remove the singularity
at $\zeta_j$ to produce a multiplet regular in $\cU_j$, and since
$\log\eta\sim\log(1-\ztp/\zeta)+\log(\zeta-\ztm)$, one concludes that in case (i)
one must have $\cij{0\infty}=\cij{0-}$, whereas in the second case one
must have $\cij{0+}=-\cij{0-}$.\footnote{Actually, these two cases are related by a
gauge transformation \eqref{gaugeHG}. See the example of the improved tensor multiplet
in Section \ref{subsec_secondflat}.}

\lfig{Structure of branch cuts  in case (i), $\cij{0-}=\cij{0\infty}$. The solid (black) line represents the
branch cut in $\muor(\zeta)$, the semi-dotted (red) line is the branch cut in $\Hij{0-}$, and the
dotted (blue) line is the contour $C$ along which $\muor(\zeta)$ is integrated.}{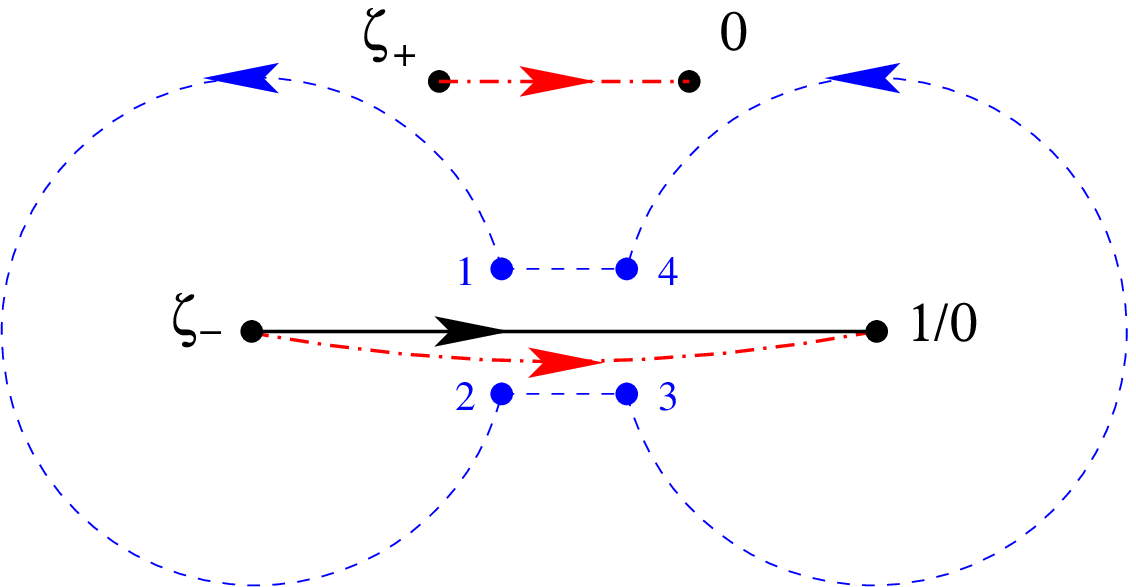}{5cm}{figure0}

\lfig{Structure of branch cuts  in case (ii), $\cij{0+}=-\cij{0-}$. The solid line represents the
branch cut in $\muor(\zeta)$, and the semi-dotted line is the branch cut in $\Hij{0+}$.
On the left, the dotted line is the contour along which $\muor(\zeta)$ is integrated. On the
right, the dotted line is the contour along which the singular part $c^{[0+]} \log \eta$ of $\Hij{0+}$
is integrated. The part of the contour between 2 and 4 lies on the $n+1$-th Riemann sheet of
the logarithm. }{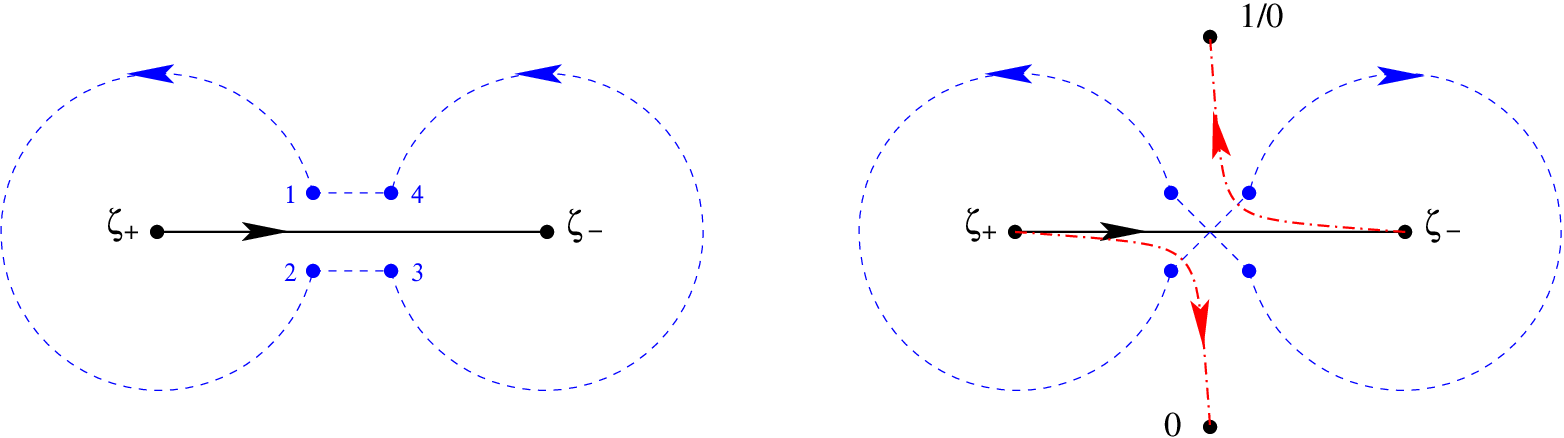}{4.2cm}{figure8}

Let us consider these two possibilities in some more detail. The first situation is shown on Fig. \ref{figure0}.
The branch cut in $\Hij{0-}$ can be chosen to run from $\zeta_-$ to $\infty$ and $\zeta_+$ to $0$.
 Since we integrate the same function, the two open contours (12) and (34) can be joined into a  single
closed contour which encircles the cut from $\ztm$ to $\infty$.

In the second case, the contour along which  $\muor(\zeta)$ is integrated initially runs around
the cut from $\ztm$ to $\ztp$. Along the contour (12) (resp. (34)), one may replace $\muor(\zeta)$
by $\mui{+}-\pa_\eta(\Hij{0+}_{\rm reg}+\cij{0+}\log\eta)$ 
(resp. $\mui{-}-\pa_\eta(\Hij{0-}_{\rm reg}+\cij{0-}\log\eta)$).
The branch cut in $\log\eta$ can be chosen to extend  from $\ztp$ to $0$ and
 from $\ztm$ to $\infty$. The contributions from the terms $\mu\ui{\pm}$ drop by the regularity assumption, and the
relative negative sign between $\cij{0+}$ and $\cij{0-}$ can be eliminated by reversing the
contour (34). In the limit where all points 1,2,3,4 coincide, which also implies that the
two branch cuts touch each other,
it is possible to reconnect the contours (12) and (43) into a single ``figure-eight" 
closed contour, recovering the prescription in  \cite{Karlhede:1984vr}.
The cuts can then be separated as shown in Fig. \ref{figure8}, in such a way
that the part of the contour which extends from 2 to 4 crosses into the $n+1$-th Riemann sheet
of the  logarithm.

The upshot of this discussion is that in cases with logarithmic branch cuts,
the same formulae \eqref{tlag1}, \eqref{mui}
as in the  meromorphic case apply, with the proviso
that the integration contours do not just encircle the patches,
but encircle or trace a figure-eight contour around the logarithmic branch cuts,
depending on the signs of the logarithmic terms $\cij{ij}\log\eta$ in the
transition functions $\Hij{ij}$.

\subsection{Example:  $\IR^4$}

We now illustrate the general formalism developed in this section in the simplest case
of flat four-dimensional Euclidean space $\IR^4$. We present two equivalent
descriptions, each with its own virtues.

\subsubsection{Free tensor multiplet}
\label{subsec_firstflat}

The simplest description of the twistor space of  $\IR^4$ involves a single $\cO(2)$
multiplet and two patches around $\zeta=0$ and $\zeta=\infty$, with transition function
\be
\Hij{0\infty}=\frac12\,\eta^2 \, .
\ee
This is referred to as the ``free tensor multiplet'' in the physics literature.
In order to facilitate the comparison with later subsections, we deviate
from the notation in \eqref{omult} and set
\be
\eta = z/\zeta + y - \bz \zeta\, .
\ee
The tensor Lagrangian \eqref{tlag1} evaluates to
\be
\mathcal{L}(y,z,\bar z) =
-\frac12\, y^2 + z \bar z \, .
\label{lflatt}
\ee
Performing the Legendre transform with respect to $y$ gives a K\"ahler potential
\be
K = \frac12 ( w+ \bar w)^2 + z \bar z\ ,\qquad w+\bar w = - y
\label{kflatt}
\ee
for the flat metric $\de s^2=\de z \de\bz+\de w \de\bw$ in the complex structure $J^3$.
Evaluating \eqref{solmu} and \eqref{solmub}, we obtain
\be
\label{mu0inf4}
\muor = w + \bar z \zeta\, , \quad \muinf  = -\bar w + z/\zeta
\ee
consistently with the requirements
\be
\muinf - \muor =\p_\eta \Hij{0\infty} \, ,
\qquad
\muinf = - \overline{\tau(\muor)} \, .
\ee
Moreover, the holomorphic section $\Omega$ reproduces the correct
\kahler forms
\be
\label{eq:1.26}
\Omega = \zeta \de \muor \wedge \de\eta =
\de w\wedge \de z + \zeta ( \de\bz \wedge \de z +\de \bw \wedge \de w)
+ \zeta^2 \, \de \bw\wedge \de \bz \, .
\ee
Observe that $\I \zeta \eta$ is the $\cO(2)$ valued moment map for the tri-holomorphic
isometry $V=i(\pa_w - \pa_{\bar w})$, in the sense that $\mathcal{L}_V \Omega = \de (\I \zeta\eta)$.

We now relate this to the standard construction of the twistor space of $\IR^4$,
in \cite{Penrose:1972ia}. First, observe that despite their linear form,
$\muor$ and $\muinf$ do {\em not} patch together into a single
$\cO(1)$ global section. Rather, the doublet $(\mu_1\ui{0},\mu_1\ui{\infty})=(\muor,\muor/\ze)$,
$(\mu_2\ui{0},\mu_2\ui{\infty})=(\zeta\muinf,\muinf)$ define two different
global sections of $\cO(1)$,
related by a symplectic (Majorana) reality condition
$\muor_A =  \epsilon_{AB} \overline{\tau(\muinf_B)}$.
Note that Eqs. \eqref{mu0inf4} may be rewritten as
\be
\label{muzpi}
\begin{pmatrix} \muor \\ \ze \muinf \end{pmatrix} =
\begin{pmatrix} \bz & w \\ - \bw & z \end{pmatrix}\cdot
\begin{pmatrix} \zeta \\ 1 \end{pmatrix} \, .
\ee
Defining the doublet $\pi^{A'}_{[0]}=(\zeta,1), \pi^{A'}_{[\infty]}=(1,1/\zeta)$,
and $x_{AA'}={\scriptsize \begin{pmatrix} \bz & w \\ - \bw & z \end{pmatrix}}$,
Eq. \eqref{muzpi} becomes
\be
\mu_{A}^{[i]} = x_{AA'} \pi^{A'}_{[i]} \ , \quad i=0,\infty\, .
\ee
This reproduces the standard relation between points in $\IR^4$ and lines in $\IC P^3$,
or two-planes in $\IC^4$, coordinatized by $(\mu_A, \pi^{A'})$.
Unlike the construction in the next subsection, it should be noted that
the K\"ahler potential \eqref{kflatt} is {\it not} invariant
under the $SU(2)$ action by right multiplication on the matrix $x_{AA'}$.

\subsubsection{Improved tensor multiplet}
\label{subsec_secondflat}

We now describe an equivalent construction of the twistor space of $\IR^4$,
known as the ``improved tensor multiplet'' in the physics literature \cite{deWit:1981fh,Karlhede:1984vr}, 
which makes the superconformal symmetry manifest (see \cite{Alexandrov:2008nk} for a
discussion of superconformal properties).

Let us introduce four different patches $\cU_0,\ \cU_\pm$, $\cU_{\infty}$ and
the following transition functions
\be
\Hij{0+}= -\frac12 \eta' \log \eta' \ ,\qquad
\Hij{0-}= \frac12 \eta' \log \eta' \ ,\qquad
\Hij{0\infty}=0\, ,
\label{trfflat}
\ee
where $\eta'(\zeta)=v/\ze+x-\vb \ze$ is a global section of $\cO(2)$
(denoted by a prime to avoid confusion with the previous construction).
This set of transition functions falls into case (ii) considered
in Subsection \ref{subsec_branch}.
The Lagrangian for this system is given by the 
contour integral representation \cite{Karlhede:1984vr}
\be
\label{Limpro}
\mathcal{L}= -\frac12 \oint_{C} \frac{\de\zeta}{2\pi \I\,\zeta} \eta' \log \eta' \, ,
\ee
where $C$ is the ``figure-eight" contour, encircling the roots $\ztp$ and $\ztm$ of
$\eta'$,
\be
\ztpm=\frac{x\mp r}{2\bv}\, ,\qquad
\ztp \overline{\ztm} = -1\, ,\quad \ztp \ztm = -v/\bv\ ,
\label{etazeros}
\ee
counterclockwise and clockwise, respectively, and the radial variable $r$ is
defined as
\be\label{def-r}
r^2= x^2+4v\bar v\ .
\ee
The contour integral \eqref{Limpro} leads to
\be
\label{lflat2}
\mathcal{L}(x,v,\bar v) = r - x \log\frac{x+ r}{2|v|}\, .
\ee
Note that $\cL$ is invariant under phase rotations $v\to e^{i\theta} v$, and is
homogeneous of degree one in $x,v,\bv$. Dualizing $x$ into $u+\bu=\pa_x \cL$,
we find
\be
K = 2|v|\, \cosh(u+\bar u)\, ,\qquad
x = -2|v|\, \sinh(u+\bar u)\, .
\ee
Changing variables to
\be\label{CTFS2}
u=\frac12\,\log \frac{w}{z}\, ,\qquad
v= z w  \, ,
\ee
leads to the flat space K\"ahler potential
\be
K=z \bz+ w \wb\ . \label{kinv}
\ee
Moreover,
\be
\label{etaprime}
\eta' = z w/\zeta +( z\bar z- w \bar w)  - \bar z \bar w \zeta\ ,
\ee
is identified (up to a factor $\I \zeta$) as the $\cO(2)$-valued moment map
associated to the rotation $V'=i( w \pa_w - z \pa_z - \bar w \pa_{\bar w} +\bar z \pa_{\bar z})$.

The twistor line $\mu'^{[0]} $ may be computed from \eqref{solmu} using the same contour $C$,
leading to
\be
\label{muFS2}
\mu'^{[0]} =
\frac{\I}{2} \vrh - \frac12 \log\frac{x+ \sqrt{x^2+4 v \bar v}}{2|v|}
+\frac12\,\log\frac{1-\zeta/\zeta_+}{1-\zeta/\zeta_-}
=\frac12\, \log\frac{w+\bar z \zeta}{z-\bar w \zeta} \, .
\ee
The real structure maps $\mu'^{[0]}$ to itself, in agreement with the fact
that $\Hij{0\infty}=0$.  While  $\mu'^{[0]} $ is regular around $\zeta=0$ (as it should),
it has a logarithmic branch cut on the segment $[\zeta_-,\zeta_+]=[z/\bw,-w/\bz]$.
The logarithmic singularity at $\ze=\ze_+$ cancels from the multiplet
 \be
\mu'^{[+]} = \mu'^{[0]} - \frac12 \left(\log \eta' +1 \right)\, ,
\ee
at the expense of introducing logarithmic singularities at $\zeta_-, 0$ and $\infty$.
Similarly, $\mu'^{[-]} = \mu'^{[0]} + \frac12 (\log \eta'+1)$ is regular at $\zeta_-$
but has a logarithmic singularities at $\zeta_+$, 0 and $\infty$. Altogether,
$\mu'^{[0]}$ and $\mu'^{[\pm]}$  along with $\eta'$ provide regular Darboux
coordinates throughout the $\zeta$ plane. They are related to 
the  global  $\cO(1)$ sections $\mu_A$ defined above \eqref{muzpi} by 
\be
\mu_1^{[0]} = \sqrt{\zeta}  \exp\left( \mu'^{[-]} -\frac12 \right)\ ,\quad 
\mu_2^{[0]} = \sqrt{\zeta} \exp\left( - \mu'^{[+]} -\frac12\right)\ ,\quad 
\ee
whose product  is the  global $\cO(2)$ section $\zeta \eta'=\mu_1^{[0]} \mu_2^{[0]}$.
This recovers the standard construction of the twistor space of $\IR^4$ as the
spectral curve $x(\zeta) y(\zeta)=z(\zeta)$ \cite{besse1987em}.
The relation between the complex Darboux coordinates $\eta', \mu'$
and  $\eta, \mu$ in the patch $\cU_0$ is given by the  symplectomorphism
generated by
\be
S(\eta,\mu') = \frac12 \frac{\zeta  \eta^2e^{2\mu'} }{1-\zeta e^{2\mu'}}=\frac12 \eta \mu\,  ,
\qquad
\eta' = \mu\,(\mu+\eta)\, ,
\quad
\mu' = \frac12\,\log \frac{\mu}{\zeta (\mu+\eta)}   \, .
\ee
In contrast to \eqref{kflatt}, the K\"ahler potential is invariant under the $SU(2)$ action,
and is equal to the \hk potential of $\IR^4$. As we discuss further in \cite{Alexandrov:2008nk}, the
$SU(2)$ symmetry is realized by fractional linear transformations of $\zeta$.
Indeed it is easy to check that
\bea
\left(\wb \pa_z - \zb \pa_w + \pa_\zeta \right) \nu &=& 0\ ,\nn\\
\left( z\pa_z - \zb\pa_{\zb} + w \pa_w - \wb\pa_{\wb} + 2 \zeta \pa_\zeta
-2n \right)  \nu &=&0\ , \\
\left( w \pa_{\zb} - z \pa_{\wb} + \zeta^2 \pa_\zeta-2n\zeta\right) \nu&=& 0\ ,\nn
\eea
where $(\nu,n)=(\zeta \eta',1)$ and $(\mu',0)$, respectively.

Finally, we mention that an equivalent description may be obtained by
applying the gauge transformation \eqref{gaugeHG} to the transition
functions \eqref{trfflat}, with
\be
\Gi{0}=\hf\eta'\log(\eta'\zeta)\, ,\quad
\Gi{\infty} = -\hf\eta'\log(\eta'/\zeta)\ ,\quad
\Gi{+}=\Gi{-}=\frac12 \eta' \log\zeta\, \ .
\ee
This results into a new set of transition functions with
\be
\Hij{0+}=0\, ,
\qquad
\Hij{0-}=\Hij{0\infty}=\eta'\log\eta'\,,
\ee
representative of case (i) in Section \ref{subsec_branch}. The Lagrangian now reads \cite{Alexandrov:2007ec}
\be
\label{lflat3}
\CL=\oint_{C} \frac{\de \zeta}{2\pi \I\, \zeta}\, \Hij{0\infty}= \sqrt{x^2 + 4 v \bar v}
-x-x\log\frac{x+ \sqrt{x^2 + 4 v \bar v}}{2}\, ,
\ee
where the contour $C$  encircles the logarithmic branch cut from $\zeta_-$ to $\infty$.
While \eqref{lflat3} is invariant under a phase rotation of $v$, unlike \eqref{lflat2} it is not
a homogeneous function of $x,v,\bv$. Rather, it satisfies
\be
\left( x \pa_x +  v \pa_v +  \vb \pa_{\vb} - 1\right) \cL = -  x\, .
\ee
The occurrence of a linear term in $x$ on the r.h.s. of this equation is
consistent with superconformal invariance, we shall discuss it
further in \cite{Alexandrov:2008nk}.
The twistor line $\mu'^{[0]} $ following from \eqref{mui} is now given by
\be
\label{mu0mt}
\mu'^{[0]}=\frac{\I}{2}\, \vrh'-\log\(1-\frac{\zeta}{\ztm}\)-\hf\(\log\frac{x+r}{2}+1\)\, .
\ee
In agreement with the general discussion in section \ref{subsec_branch},
$\mu'^{[0]}$ has logarithmic singularities at $\ztm$ and $\infty$. It can be checked
that $\vrh'$ in this equation differs from $\vrh$ in \eqref{muFS2} by
$\vrh'=\vrh+(1/2i) \log(v/\bv)$.

\section{Linear deformations of $\cO(2)$ hyperk\"ahler spaces}
\label{sect:lindef}

In this section, we finally discuss the infinitesimal deformations of $\cO(2)$ manifolds (i.e.
$4d$-dimensional HK manifolds with $d$ commuting tri-holomorphic isometries) which
preserve the \hk property but may break all isometries. Our strategy is to perturb
the functions $\Sij{ij}$ in \eqref{trans_2n} generating the symplectomorphisms
between two patches by arbitrary functions of $\nui{i}^I, \mui{i}_I, \ze\ui{i}$, in a
way compatible with the consistency constraints \eqref{consist_cond} and
reality conditions \eqref{Sreal}, and work out the corrections to first order in the
perturbation. Note that we do not perturb the transition functions $f_{ij}$ of
the $\cO(1)$ line bundle on $\CP$, nor the real structure, which can always
be chosen as in  \eqref{realcon} by use of a local symplectomorphism.
The same strategy could also be applied to $\cO(2n)$ manifolds with $n\geq 2$,
but in view of practical applications we restrict to the case $n=1$.

\subsection{Deforming the symplectomorphisms}

In line with the strategy just outlined, we consider deformations of
the generating functions \eqref{trans_2n} of the following form
\be
\Sij{ij}(\nui{i},\mui{j},\zeta\ui{i})=f_{ij}^{-2}\nui{i}^I\mui{j}_I
-\tHij{ij}(\nui{i},\zeta\ui{i})-\tHpij{ij}(\nui{i},\mui{j},\zeta\ui{i}) \, ,
\label{genf}
\ee
where $\tHpij{ij}(\nui{i},\mui{j},\zeta\ui{i})$ are functions of $2d+1$ variables
subject to the compatibility constraints to be discussed presently. These
functions generate symplectomorphisms
\be
\nui{j}^I = f_{ij}^{-2}\nui{i}^I -\p_{\mui{j}_I}\tHpij{ij} \, , \qquad
\mui{i}_I =\mui{j}_I-f_{ij}^2\(\p_{\nui{i}^I}\tHij{ij}+\p_{\nui{i}^I}\tHpij{ij}\) \, .
\label{ourcantr}
\ee
Our task is to determine the linear corrections $\nupi{i}^I,\mupi{i}_I $
to the unperturbed twistor lines $\nuzi{i}^I,\muzi{i}_I $ such that
\be
\nui{i}^I=\nuzi{i}^I +\nupi{i}^I \, , \qquad
\mui{i}_I=\muzi{i}_I+\mupi{i}_I\ ,
\ee
 satisfy \eqref{ourcantr}, to linear order in $\tHpij{ij}$. Here $\nuzi{i}^I,\muzi{i}_I $
 are given by $\nuzi{i}^I=\zeta f_{0i}^{-2}\eta^I$ and \eqref{mui}, respectively.

As in \eqref{HHt}, it is advantageous to express $\tHpij{ij}$ in terms of the unperturbed
$\cO(2)$ global sections $\eta^I$ and the local coordinate $\zeta=\zeta\ui{0}$ around
the north pole, and define
 \be
 \Hpij{ij}(\eta,\muzi{j},\zeta) \equiv \zeta^{-1}f_{0j}^2  \,\tHpij{ij}(\zeta f_{0i}^{-2}\eta,\muzi{j},\zeta)\, .
 \ee
 The symplectomorphisms \eqref{ourcantr} are then rewritten as
\bse
\label{smalleq}
\bea
\label{smalleq1}
\nupi{j}^I&=&f_{ij}^{-2}\nupi{i}^I-\zeta f_{0j}^{-2}\,\p_{\muzi{j}_I}\Hpij{ij} \ ,\\
\mupi{j}_I&=&\mupi{i}_I+\p_{\eta^I}\Hpij{ij}+\zeta^{-1}f_{0i}^2\,\nupi{i}^J\p_{\eta^I}\p_{\eta^J}\Hij{ij} \, ,
\label{smalleq2}
\eea
\ese
while the consistency \eqref{consist_inv}, \eqref{consist_cond} and reality conditions \eqref{Sreal} become
\be
\Hpij{ji}(\eta,\muzi{i},\zeta)=-\Hpij{ij}(\eta,\muzi{j},\zeta)\ ,
\ee
together with
\be
\Hpij{ij}(\eta,\muzi{j},\zeta)=\Hpij{ik}(\eta,\muzi{k},\zeta)+\Hpij{kj}(\eta,\muzi{j},\zeta)\ ,
\label{consist_condHp}
\ee
and
\be
\overline{\tau(\Hpij{ij})}=-\Hpij{\bi\bj}\, ,
\label{realconH}
\ee
respectively.
In these expressions, the arguments $\muzi{j}$ on various patches are to be
related
to each other using the unperturbed equations \eqref{eq22}. These conditions can be further simplified by expressing all
$\muzi{j}$ in terms of a single real multiplet\footnote{Of course, similar multiplets
$\rhoi{i}_I(\zeta)\equiv -\I\left(\muzi{i}_I + \muzi{\bi}_I \right)$ can be introduced in each
patch; we focus on the one relevant for the complex structure $J^3$ at $\ze=0$.}
\be
\rho_I\equiv -\I\left(\muzi{0}_I + \muzi{\infty}_I \right)
=-\I\left(2\muzi{i}_I+ \p_{\eta^I} (\Hij{i0}+\Hij{i\infty})\right) \, .
\label{newvar}
\ee
Using \eqref{mui}, $\rho_I$ can be written in the following form
\be
\rho_I(\zeta)= \vrh_I-\I\sum_j \oint_{C_j} \frac{\de\zeta'}{2\pi \I\, \zeta'}
\, \frac{\zeta'+\zeta}{\zeta'-\zeta}\,\p_{\eta^I}\Hij{ij}(\zeta')
-i\p_{\eta^I} (\Hij{i0}+\Hij{i\infty}) \, ,
\label{newvar2}
\ee
where the pole at $\zeta'=\zeta$ is as usual inside $C_i$.
Similarly to $\eta^I$, the multiplet $\rho_I$ is manifestly real, $\overline{\tau(\rho_I)}=\rho_I$,
so the reality conditions \eqref{realconH} are automatically obeyed if one requires that
$\Hpij{i\bi}$ is a real function of $\eta^I$ and $\rho_I$.

\subsection{Deformed twistor lines}

We now proceed to determine the perturbations of the twistor lines,
solving \eqref{smalleq1} for $\nupi{i}^I$, substituting the result in
 \eqref{smalleq2}, and finally determine $\mupi{i}_I$.  The procedure
 follows the same steps as in Section \ref{seco2n}, and we only
 quote the result. Before this, note however that \eqref{smalleq}
 define $\nupi{i}^I$ and $\mupi{i}_I$ only up to the addition of
  global $\cO(2)$ and $\cO(0)$ sections, respectively.
 Consistently with the reality constraints \eqref{realcon},
 we may fix these ambiguities by imposing
\be
\bar\hnu_{[0]1}^I=-\hnu_{[0]1}^I\, ,
\qquad
\bar\hnu_{[0]0}^I = \hnu_{[0]2}^I\, ,
\qquad
\bar \hmu^{[0]}_{I,0}=\mupi{0}_{I,0} \, .
\ee
In this way, we obtain
\be
\begin{split}
\nupi{i}^I =&\, \I f_{0i}^{-2}\sum_j\oint_{C_j} \frac{\de\zeta'}{2\pi \I\,\zeta'}\,
\frac{\zeta^3+\zeta'^3}{\zeta' (\zeta'-\zeta)}\, \Hp^{[0j]I}(\zeta') \, ,
\label{nupi}\\
\mupi{i}_I =&\, \sum_j\oint_{C_j} \frac{\de\zeta'}{2\pi \I\,\zeta'}\, \frac{\zeta+\zeta'}{2 (\zeta'-\zeta)}\, \Gi{0j}_I(\zeta') \, ,
\end{split}
\ee
where
\be
\label{defgi}
\Gi{ij}_I \equiv  {\Hpij{ij}}_I+\zeta^{-1}f_{0i}^2\nupi{i}^J\Hij{ij}_{IJ}
+\I \Hp^{[ij]J}(\Hij{j0}_{IJ}+\Hij{j\infty}_{IJ})\ ,
\ee
and, as usual, $\zeta$ is restricted to lie inside the contour $C_i$.
Here and below, we use the shorthand notation
\be
H_I\equiv \p_{\eta^I}H\, ,
\qquad
H_{IJ}\equiv \p_{\eta^I}\p_{\eta^J} H\, ,
\qquad
{\Hp}_I\equiv \p_{\eta^I}\Hp  \, ,
\qquad
\Hp^I\equiv \p_{\rho_I}\Hp\, .
\ee
The second term in \eqref{defgi} corresponds to the variation of the unperturbed 
$\muzi{i}_I$ under $\nuzi{i}^I  \to \nuzi{i}^I + \nupi{i}^I$, while the third term is due
to the change of variable from $(\nui{i}^I,\mui{i}_I)$ to $(\eta^I,\rho_I)$.
It is easy to verify that $\overline{\tau[\Gi{ij}_I]}=-\Gi{\bi\bj}_I$, in accordance with
the reality conditions \eqref{realcon}.

\subsection{Perturbed \kahler potential and Penrose transform}

Having obtained the deformed twistor lines, the holomorphic 2-form
$\omega^+$ and the K\"ahler form $\omega^3$
in the complex structure $J^3$ can be obtained as usual by Taylor expanding
the holomorphic section $\Omega\ui{0}=\de\mui{0}_I \wedge \de\nui{0}^I$ around $\zeta=0$.
The constant term gives
\be
\label{omplus}
\omega^+
=\de\(\frac{\I}{2}\, \vrh_I+\hf\oint_C \frac{\de\zeta}{2\pi \I\, \zeta}\(H_I+G_I\) \)
\wedge \de\(v^I+\I \oint_C \frac{\de\zeta}{2\pi \I}\, \Hp^I \)  \, ,
\ee
where, here and henceforth, we suppress the patch indices and denote
the sum over contours $C_i$ as a single contour $C$.
Eq. \eqref{omplus} identifies
\be
u^I \equiv   v^I+\I \oint_C \frac{\de\zeta}{2\pi \I}\, \Hp^I  \, ,
\qquad
w_I \equiv  \frac{\I}{2}\, \vrh_I+\hf\oint_C \frac{\de\zeta}{2\pi \I\, \zeta}\(H_I+G_I\)
\ee
as holomorphic Darboux coordinates on $\cM$.
Using \eqref{newvar2} and \eqref{intF1},
they can be rewritten as
\be
u^I =   v^I+\I\p_{\vrh_I} \left( \oint_C \frac{\de\zeta}{2\pi \I}\, \Hp \right) \, ,
\qquad
w_I = \hf\(y_I + \I\vrh_I\) \, ,
\label{holcoor}
\ee
where we introduced the real quantity
\be
\begin{split}
y_I =&\p_{x^I} \left( \oint_C \frac{\de\zeta}{2\pi \I\, \zeta}\(H+\Hp\) \right)
\label{Brho}\\&
-\I \(\oint_C \frac{\de\zeta}{2\pi \I\, \zeta^2}\, \Hp^J \oint_C \frac{\de\zeta}{2\pi \I}\,H_{IJ}
- \oint_C \frac{\de\zeta}{2\pi \I}\, \Hp^J \oint_C \frac{\de\zeta}{2\pi \I\, \zeta^2}\,H_{IJ} \) \, .
\end{split}
\ee
On the other hand, the $\cO(\zeta)$ term in $\Omega\ui{0}$ leads to the \kahler form
\be
\label{Kform}
\omega
=\I \[\de\(\oint_C \frac{\de\zeta}{2\pi \I\, \zeta^2}\(H_I+G_I\)\)\wedge
\de u^I + \de w_I
\wedge
\de\(x^I+\I \oint_C \frac{\de\zeta}{2\pi \I\,\zeta}\, \Hp^I \)\] \, .
\ee
The \kahler potential $K(u,\bu,w,\bw)$ therefore should satisfy
\be
\label{dK}
K_{u^I} = \oint_C \frac{\de\zeta}{2\pi \I\, \zeta^2}\(H_I+G_I\)  \, ,
\qquad
K_{w_I} = - x^I -\I \oint_C \frac{\de\zeta}{2\pi \I\,\zeta}\, \Hp^I  \, .
\ee

To integrate these equations, it is convenient to introduce the ``perturbed Lagrangian"
\be
\label{L01}
\CL=\oint_C \frac{\de\zeta}{2\pi \I\, \zeta}\(H +\Hp\)\equiv \CL_{(0)} + \CL_{(1)} \ ,
\ee
a function of $v^I,\bv^I,x^I,\vrh^I$; the derivatives of $\CL$ will be denoted
by $\CL_{x^I}=\p_{x^I}\CL$, $\CL_{v^I}=\p_{v^I}\CL$, etc.
It is also convenient to introduce a special notation for the coefficients
\be
\nupk{0}^I\equiv \hat\nu^I_{[0]0}= \oint_C \frac{\de\zeta}{2\pi }\, \Hp^I \, ,
\qquad
\bnupk{0}^I\equiv \bar{\hat\nu}^I_{[0]0}= \oint_C \frac{\de \zeta}{2\pi \zeta^2}\, \Hp^I \, .
\ee
Using \eqref{intF2} in Appendix \ref{ap_integr}, one can rewrite
the complex coordinates  as
\be
u^I =   v^I+\nupk{0}^I\, ,
\qquad
w_I = \frac12 \left( \I \vrh_I +
\CL_{x^I}+\nupk{0}^J\CL_{v^Jx^I}+\bnupk{0}^J\CL_{\bar v^Jx^I} \right)
\ee
and \eqref{dK} as
\be
K_{u^I} = \CL_{v^I}+\nupk{0}^J\CL_{v^Jv^I}+\bnupk{0}^J\CL_{\bv^Jv^I}\, ,
\qquad
K_{w_I} = - \(x^I+\I\CL_{\vrh_I}\)   \, .
\label{derchi}
\ee
These two equations can be integrated into
\be
K(u,\bu,w,\bw)=\langle \CL(v,\bv,x,\vrh)+\nupk{0}^I\CL_{v^I}+\bnupk{0}^I\CL_{\bar v^I}-x^I y_I\rangle_{x^I} \ .
\ee
Since $u^I =   v^I+\nupk{0}^I$, one may rewrite this more concisely as
\be
\label{Kdef}
K(u,\bu,w,\bw)= \langle \CL(u,\bu,x,\vrh) -x^I (w_I+\bw_I) \rangle_{x^I}\, .
\ee
Thus, a \kahler potential $K$ for the perturbed HK metric
may be obtained by Legendre transform from the perturbed Lagrangian
$\CL$ defined in \eqref{L01}, now considered as a function
of $u^I,\bu^I,x^I$ and $\vrh_I=-\I(w_I-\bw_I)$. In particular,
the variation of the \kahler potential
is given to first order in the perturbation by
\be
\label{kahlerdef}
K_{(1)}(u,\bu,w,\bw)=
\sum_j \oint_{C_j} \frac{\de\zeta}{2\pi \I\,\zeta}\,
\Hpij{0j}( \eta, \rho,\zeta) \, ,
\ee
where $K_{(1)}(u,\bu,w,\bw)\equiv K(u,\bu,w,\bw)-K_{(0)}(u,\bu,w,\bw)$,
and $K_{(0)}$ is given by the Legendre transform \eqref{K02}
of the unperturbed Lagrangian,
 \be
K_{(0)}(u, \bu,w, \bw) = \langle \cL_{(0)}(u^I,\bu^I,x^I) - x^I  (w_I + \bw_I) \rangle_{x^I} \ .
\label{K022}
\ee
while $\eta^I$ and $\rho_I$ in \eqref{kahlerdef} are the unperturbed $\cO(2)$ and
conjugated multiplet \eqref{newvar2}, evaluated at $(x^I,u^I,\bu^I)$.

Equation \eqref{kahlerdef} is one of the main results of this paper.
In words, it says that the variation of the K\"ahler potential is given by a Penrose-type
contour integral of the holomorphic section $\Hp(\eta,\rho,\zeta)$, along the fibers
of the projection $\pi:\cZ\to\cM$. It is consistent with the fact that  $K_{(1)}$
should be a zero-eigenmode of the Laplace-Beltrami operator on $\cM$, as required by the
linearization of the Monge-Amp\`ere equation.

As usual, the perturbed metric may be computed without
knowing $x^I$ as a function of $u^I, \bu^I,w_I, \bw_I$ explicitly, using
a generalization of \eqref{HKmet0},
\be\label{HKmet}
\begin{split}
K_{u^I \bu^J} =& \, \CL_{u^I \bu^J}-\CL_{u^I x^K}\CL^{x^K x^L}\CL_{x^L \bu^J} \, ,
\\
K_{u^I \bw_J} =& \, \CL_{u^I x^K}\CL^{x^K x^J}+\I\(\CL_{u^I\vrh_J}-\CL_{u^I x^K}\CL^{x^K x^L}\CL_{x^L \vrh_J}\) \, ,
\\
K_{w_I \bu^J} =& \, \CL^{x^I x^K}\CL_{x^K \bu^J}-\I\(\CL_{\vrh_I \bu^J}-\CL_{\vrh_I x^K}\CL^{x^K x^L}\CL_{x^L \bu^J}\) \, ,
\\
K_{w_I \bw_J} =& \, -\CL^{x^I x^J}+\CL_{\vrh_I\vrh_J}+\I\(\CL^{x^I x^K}\CL_{x^K \vrh_J}-\CL_{\vrh_I x^K}\CL^{x^K x^J}\) \, .
\end{split}
\ee
In Appendix  \ref{ap_integr}, we check directly that the deformed metric is  \hk, and provided
formulae for the inverse metric.

\section{Examples}

We now illustrate our general formalism on two well-known examples,
namely Taub-NUT space, which we express as a deformation of
flat $\IR^4$ (preserving one tri-holomorphic isometry), and the Atiyah-Hitchin
manifold, which we  express as a deformation of Taub-NUT space (breaking
all tri-holomorphic isometries).

\subsection{Twistorial description of Taub-NUT space} \label{subsect:TN}

The Taub-Nut manifold is described by the superposition of the two Lagrangians \eqref{lflatt}
and \eqref{lflat2} describing flat $\IR^4$ 
\cite{Karlhede:1984vr,Hitchin:1986ea},
\bea
\cL &=& - \frac{1}{2} \, \oint_{C} \frac{\de \zeta}{2 \pi \I \zeta} \, \eta \, \ln \eta
 - c  \, \oint_{C_0} \frac{\de \zeta}{2 \pi \I \zeta} \, \eta^2   \nn\\
 &=& \big( r - x \ln(r+x) + \tfrac{1}{2} x \ln(4 v \vb) \big) - 2 c \big( - \tfrac{1}{2} x^2 + v \vb \big) \, ,
 \label{TN1}
\eea
where $c$ is related to the mass parameter as $c = -1/m$.
The Legendre transform of $\cL$ leads to the K\"ahler potential
\be
\label{TN3}
K = \cL - x (u + \bar u) =r - c ( x^2 + 2 v \bar v)  \ ,
\ee
where $x$ is given in terms of $v,\bv,u+\bu$ by the transcendental equation
\be
\label{TN2}
u + \bar u  = 2 c x -{\rm arcsinh}\left(x/\sqrt{4 v \vb}\right)\ .
\ee
The derivatives of $K$ can be computed without
solving equation \eqref{TN2} explicitly. Using the relations \eqref{HKmet0}
for $\cL$ being $\varrho$-independent one finds
\be\label{TN4}
\begin{split}
K_{u\bar u}
= \frac{r}{1-2 c r}\ ,\quad
K_{u\bar v}=
- \frac{x}{2\vb(1-2 c r)}\ ,\quad
K_{v\bar v}=
\frac{1-2 c r}{r}+\frac{x^2}{4 v\vb\, r (1-2 c r)}\ .
\end{split}
\ee
Defining spherical coordinates
\be
\label{sphtn}
x= r \cos\theta\ ,\quad v = \frac12 r\,e^{i\phi}\,\sin\theta\ ,\quad
\bv = \frac12 r\,e^{-i\phi}\,\sin\theta\ ,\quad
u -  \bar u = - i \psi \, ,
\ee
the resulting metric takes the usual form,
\be
\label{tnmetric}
4\,ds^2 = V \left( dr^2 + r^2 d\theta^2 + r^2 \sin^2\theta d\phi^2 \right)+
V^{-1} \left( d\psi + \cos\theta d\phi \right)^2 \, ,
\ee
where $V$ is a harmonic function in $\IR^3$
\be
V = \frac{1}{r}-2 c \, .
\ee

For the comparison with the Atiyah-Hitchin manifold in the next section, it is useful
to note the exact expression for the local section $\mu$,
\be
\label{exactmutn}
\mu^{[0]}(\zeta) = u - 2 c \vb \zeta + \half \ln\left[ \frac{1 - \zeta/\ztp}{1-\zeta/\ztm} \right] \, ,
\quad\mu^{[\infty]} (\zeta) =
 - \bar{u} - 2 c \frac{v}{\zeta}
+ \frac{1}{2} \ln \left[ \frac{\zeta - \ztp}{\zeta - \zeta_-} \right] \, ,
\ee
where $u, v$ are related to the ``flat space'' variables $w,z$ via the coordinate
transformation \eqref{CTFS2} and $x$ is a function of $z,w,\bar z, \bar w$ via 
Eq.\ \eqref{TN2}.
In terms of the spherical coordinates $r,\theta,\phi,\psi$, the twistor lines  become
\bea
\eta(\zeta) &=& \frac1{2\zeta} r\,e^{i\phi}\,\sin\theta + r \cos\theta -  \frac12 \zeta r\,e^{-i\phi}\,\sin\theta \nn\\
\mu^{[0]}(\zeta) &=& -\frac{\I}{2}\psi - \frac12 \log \cot\frac{\theta}{2}
+\frac12\log\frac{(\cos\theta+1)\zeta + e^{\I\phi} \sin\theta}
{(\cos\theta-1)\zeta + e^{\I\phi} \sin\theta}\ ,
\nn\\
&&+ c\,r \left( \cos\theta - e^{-\I\phi} \sin\theta\,\zeta \right) \, ,
\nn\\
\mu^{[\infty]}(\zeta) &=& -\frac{\I}{2}\psi + \frac12 \log\cot\frac{\theta}{2}
+\frac12\log\frac{ \zeta \sin\theta + e^{\I\phi} (1-\cos\theta)}
{\zeta \sin\theta-e^{\I\phi} (1+\cos\theta) }
\nn\\
&&- c\,r \left( \cos\theta + e^{\I\phi} \sin\theta / \zeta \right)\, .
\label{exactmutnth}
\eea

One may check explicitly that any function of $\eta$, $\mu^{[0]}$ and
$\mu^{[\infty]}$ is a zero eigenmode of the Laplace-Beltrami operator $\Delta$
on \eqref{tnmetric}. In particular, the Penrose-type contour integral
\be
\oint_\Gamma
\frac{\de\zeta}{2\pi\I\zeta}\, H_{(1)}\left(\eta(\zeta),\mu^{[\infty]}(\zeta),\zeta\right)\,,
\ee
produces a zero eigenmode of the Laplace-Beltrami operator on Taub-NUT space,
for any choice of function $H_{(1)}$ and contour $\Gamma$. In Eq. \eqref{HtAH} below,
we identify the function $H_{(1)}$ governing the leading exponential deviation
of  Atiyah-Hitchin space away from its Taub-NUT limit.

\subsection{Taub-NUT as a deformation of  $\IR^4$}
%
To illustrate our formalism, we exhibit the Taub-Nut space as a deformation of $\IR^4$,
both in the $c\to 0$ (i.e. short distance $r$)  regime and the $c\to\infty$ (long distance) regime.
These examples are somewhat trivial, since the perturbation does not break the tri-holomorphic
isometry.

In the limit $c\to 0$ keeping $u,v$ fixed, Eq. \eqref{TN2} can be solved to first order in $c$,
\be
x = |v| \big( \e^{-(u + \bar u)} - \e^{u + \bar u} \big)
+ 2 c |v|^2 \big( \e^{-2(u + \bar u)} - \e^{2(u + \bar u)} \big)
+ \cO(c^2) \, .
\ee
Substituting in \eqref{TN3}, this leads to
\be
K =
|v| \big( \e^{-(u + \bar u)} + \e^{u + \bar u} \big)
+ c |v|^2 \big( \e^{-2(u + \bar u)} + \e^{2(u + \bar u)} - 4 \big) + \cO(c^2) \, .
\ee
Performing the coordinate transformation \eqref{CTFS2}
this becomes
\be\label{TNpotential}
K = z \bar z + w \bar w + c \big( z^2 \bar z^2 + w^2 \bar w^2 - 4 z \bar z w \bar w \big)+ \cO(c^2)  \, .
\ee
Note that the first order variation of $K$ is equal to the Lagrangian \eqref{lflatt}, evaluated
at the point \eqref{etaprime}. This indeed agrees with \eqref{kahlerdef} for a linear perturbation
with generating function
\be
\tHp(\eta,\mu,\zeta) = - c \,\eta^2\ .
\ee
Moreover, it is easy to check that the first order corrections to the sections $\eta$ and $\mu$
are also consistent with \eqref{nupi},
\bea
\eta(\zeta) &=& \frac{wz}{\zeta} + (z \bar z - w \bar w) - \bar w \bar z \zeta
+ 2 c \left( ( z \bar z )^2 - (w \bar w )^2 \right) \, ,\nn\\
\mui{0}(\zeta) &=& \half \ln \left[ \frac{w + \bar z \zeta}{z - \bar w \zeta} \right] - c\, \zeta
\left[ 2 \bar w \bar z - \frac{(w \bar w - z \bar z)(w \bar w - z \bar z + 2  \bar w \bar z \zeta)}
{(z - \bar w \zeta)(w + \bar z \zeta)} \right]\ ,\\
\mui{\infty}(\zeta) &= & \, \half \ln \left[ \frac{w + \bar z \zeta}{z - \bar w \zeta} \right] + \frac{i \pi}{2}
- c \left[ \frac{2  w  z}{\zeta} + \frac{(w \bar w - z \bar z)( \zeta(w \bar w - z \bar z) - 2  w z )}
{(z - \bar w \zeta)(w + \bar z \zeta)} \right].\nn
\eea

In the opposite regime $c\to \infty$, corresponding to distances $r \gg 1/c$,
it is appropriate to redefine $u=-2c w, v=z, x=y$. Expanding \eqref{TN2} at large
$c$ with fixed $z,w$, one obtains
\be
y = - (w+\bw) - \frac{1}{2c} {\rm arcsinh} \left( \frac{w+\bw}{2|z|} \right) +\cO(c^{-2})\ .
\ee
Substituting into \eqref{TN3} leads to
\be
K = -2 c \left[ z \bz + \frac12 (w+\bw)^2 \right]
+ \left[ \sqrt{(w+\bw)^2+4 z\bz} - (w+\bw) {\rm arcsinh} \left( \frac{w+\bw}{2|z|} \right) \right] +\cO(c^{-1})
\ee
Up to an overall factor $-2c$, the first term is equal to the \kahler potential of
flat space  \eqref{kflatt}, while the first order correction reproduces the Lagrangian \eqref{lflat2}.
This is consistent with \eqref{kahlerdef} , for a linear perturbation with generating function
\be
\tHp(\eta,\mu,\zeta) = -\frac12 \eta \log \eta\ ,
\ee
integrated along the standard ``figure-eight'' contour.

\subsection{Atiyah-Hitchin as a deformation of Taub-NUT }
\label{deftnah}

The $\cO(4)$ tensor multiplet formulation of the Atiyah-Hitchin (AH) space was
given in \cite{Ivanov:1995cy} and further analyzed in \cite{Ionas:2007gd}, to which we refer for additional details.
In this section, we compute the leading exponential correction away from
the Taub-NUT limit, and cast it into our formalism.

\subsubsection{Atiyah-Hitchin as an $\cO(4)$ manifold}
As shown in \cite{Ivanov:1995cy,Ionas:2007gd},
the AH  manifold can be obtained by the generalized Legendre transform method
from the Lagrangian
\be
\CL = 2\om\, \oint_{\Gamma_0} \frac{\de\zeta}{2\pi\I \zeta} \eta^{(4)} - \oint_\Gamma \frac{\de\zeta}{\zeta}
\sqrt{ \eta^{(4)}}\ ,
\ee
where $\Gamma_0$ encircles $\zeta=0$ while $\Gamma$ winds once around the
branch cut between the roots $\alpha$ and $-1/\bar\beta$ of the
$\cO(4)$ tensor multiplet\footnote{Note that we change notations $x\to y,
v\to \bw, z\to \bz, x_\pm \to y_\pm, v_\pm\to w_\pm,r'\to r$ with respect
to \cite{Ionas:2007gd}, to conform
to the notations of the present paper. The variable $v_\pm$ defined below
should not be confused with its namesake in \cite{Ionas:2007gd}.}
\be
\label{etao4}
\eta^{(4)} = \frac{z}{\zeta^2} + \frac{w}{\zeta}
+y - \bw \zeta +\bz \zeta^2
= \frac{\rho}{\zeta^2}\,
\frac{(\zeta-\alpha)(\bar\alpha\zeta+1)}{1+\alpha\bar\alpha}\,
\frac{(\zeta-\beta)(\bar\beta\zeta+1)}{1+\beta\bar\beta}\,\ .
\ee
In this subsection we recall how the coefficients of $\eta^{(4)}$
as well as the K\"ahler potential can be expressed in terms of elliptic functions,
which will provide a convenient starting point for extracting the Taub-NUT limit.

The section $\eta^{(4)} $ defines a real algebraic elliptic curve, given by a two-sheeted
covering of the $\zeta$ plane with branch points as $\zeta\in
\{\alpha,-1/\bar\alpha,\beta,-1/\bar\beta\}$. By a change of coordinates, the curve can be brought
into Weierstrass normal form
\be
\label{curnorm}
Y^2 = X^3 - g_2\, X - g_3\ ,
\ee
where the invariants $g_2, g_3$ are given by
\bea
\label{g23x}
g_2 &=& y_+^2 + y_+ y_- + y_-^2 +\frac14 (y_+-y_-)(w_-^2+w_+^2)\ , \\
g_3 &=& - (y_+ + y_-)\,y_+ y_- -\frac14 (y_+-y_-)(y_+ w_-^2+y_- w_+^2)\ ,
\eea
with
\be
y_\pm=\frac{y\pm 6|z|}{3}\ ,\quad w_-=\Re\frac{w}{\sqrt z}\ ,\quad w_-=\Im\frac{w}{\sqrt z}\ .
\ee
The tensor coordinates $z,w,y,\bw,\bz$
determine four points in the $(X,Y)$ plane,
\be
\label{fourpt}
\left( y_-, \pm w_- (y_+-y_-)/2 \right)\ ,\quad
\left( y_+, \pm \I w_+ (y_+-y_-)/2 \right)\ .\quad
\ee
The curve \eqref{curnorm} may be uniformized by a parameter $u\in \IC/
(2\omega\IZ+ 2\omega'\IZ)$, where $\omega$ and $\omega'$ are the half-periods,
using the Weierstrass function $\mathcal{P}(u,4g_2,4g_3)\sim 1/u^2+
\mathcal{O}(u^2)$,
\be
X = \mathcal{P}(u;4g_2,4g_3)\ ,\quad
Y = \pm \frac12 \mathcal{P}'(u;4g_2,4g_3)\ ,
\ee
where the prime denotes a $u$ derivative. The Weierstrass function
may in turn be represented in terms of Jacobi theta functions,
\be
\mathcal{P}(u;4g_2,4g_3) = \left( \frac{\pi}{2\omega} \right)^2 \left[\left(
\frac{\vth_2(0,\tau)\vth_3(0,\tau)\vth_4(v,\tau)}
{\vth_1(v,\tau)}\right)^2 - \frac{\vth_3^4(0,\tau)+\vth_2^4(0,\tau)}{3} \right]
\ ,
\ee
where the $u$ variable on the l.h.s. is related to the variable
$v$ on the r.h.s. by $v=\pi u/(2\omega)$, and 
\be
\vth_1(v,\tau)= 2\,q^{1/4}\,\sum_{n=0}^{\infty}
\,(-1)^n\,q^{n(n+1)}\,\sin\left[(2n+1)v\right]\ ,\mbox{etc}\ .
\ee
Note that we use the nome $q=e^{\pi \I \tau}$
instead of the more commonly used modular parameter $e^{2\pi \I \tau}$.
The modular parameter is $\tau=\omega'/\omega$.
We note that the half-period is set to a constant $\omega$,
which sets the scale of the Taub-NUT space, while the half-period $\omega'$,
and hence $\tau$, are purely imaginary. We abuse notation
and write $\mathcal{P}(v)=\mathcal{P}(u;4g_2,4g_3)$.

The invariants $g_2$ and $g_3$ may be expressed in terms of the
normalized Eisenstein series
$E_4=1+240 q^2+\dots$ and $E_6=1-504 q^2+\dots$ via
\be
\label{g2E4}
g_2 = \frac13 \left( \frac{\pi}{2\omega} \right)^4\,E_4\ ,\quad
g_3 = \frac2{27} \left( \frac{\pi}{2\omega} \right)^6\,E_6\ ,\quad
\ee
Having expressed 
the uniformizing variable $v=\pi u/(2\omega)$ in terms
of $X$, we define the elliptic integral of the third kind \cite{Ionas:2007gd}
\be
\pi(v) = -\frac{\pi}{2} \frac{\vth_1'(v,\tau)}{\vth_1(v,\tau)} \equiv \pi(X)\ ,
\ee
where the prime is now a $v$ derivative, and the last equality is
an abuse of notation. In particular $\pi(y_+)$ is imaginary, while
$\pi(y_-)$ is real.

The real $\cO(4)$ twistor lines are given by enforcing one condition
which expresses the $x$ coordinate in \eqref{etao4} in terms of
$z,v,\bz,\bv$. Moreover, complex coordinates
$U,Z,\bU,\bZ$ can be obtained by solving
\be
\label{UUb}
U=\pi(y_-)+\pi(y_+)\ ,\quad \bU = \pi(y_-)-\pi(y_+)\ .
\ee
The coordinates $U,Z$ are related to the coordinates $u,z$ appearing
in the Legendre transform construction by
\be
U = u \sqrt{z}\ ,\quad Z=2\sqrt{z}\ .
\ee
Here $u$ is conjugate to $v$, and should not be confused with the argument
of the Weierstrass function.

Equation \eqref{UUb}
allows to express $y_\pm$, and therefore $x$ and $|Z|$, in terms
of $U$,$\bU$ and $\tau$. In particular, we can compute the elliptic
modulus $\tau$ in terms of $U,\bU$ and $|Z|$, and plug into
the expression for $x$. The variables $v_\pm$, and therefore $v$
and $\bv$, can be computed from the fact that \eqref{fourpt} lie on
the curve. The K\"ahler potential in the complex structure
where $U,Z$ are complex coordinates is given by
\be
K = 4 \eta - (y_+ + y_-) \omega\ ,
\ee
where $\eta$ is the Weierstrass function
\be
\label{etaE2}
\eta = - \frac{\pi^2}{12\omega}\,
\frac{\vth_1^{'''}(0,\tau)}{\vth_1'(0,\tau)}
=  \frac{\pi^2}{12\omega}\,E_2(\tau)\ ,
\ee
with $E_2=1-24 q^2+\dots$ the almost-modular Eisenstein series of
weight 2.
\subsubsection{Strict Taub-NUT limit}
We now extract the limit of the Atiyah-Hitchin manifold as $\tau\to 0$.
To study this limit, we perform a modular
transformation to $\tau'=-1/\tau$. 
In the strict limit $\tau=0$, 
the elliptic curve \eqref{etao4}
factorizes as the square of a rational curve,
\be
\label{etao22}
\eta^{(4)}=\left(\eta^{(2)}\right)^2 =
\left( \frac{v}{\zeta} + x - \bv \zeta \right)^2\ .
\ee
By matching the coefficients of \eqref{etao4} and \eqref{etao22},
one readily obtains
\be
\label{zwy}
z = v^2\ ,\quad w = 2 x v \ ,\quad y = x^2 - 2 v \bv\ ,
\ee
from which it follows that
\be
\label{xpmlim}
y_+ = \frac{x^2 + 4 v \bv }{3}\ ,\quad
y_- = \frac{x^2 - 8 v \bv }{3}\ ,\quad
w_+ = 0 \ ,\quad w_- = 2 x\ .
\ee
Moreover, plugging into \eqref{g23x}, one finds
\be
g_2 = \frac13 \left( x^2 + 4 v \bv  \right)^2\ ,\quad
g_3 = -\frac2{27} \left( x^2 + 4 v \bv \right)^3\ .\quad
\ee
These last two equations are consistent with the limit $\tau\to 0$
of \eqref{g2E4} computed using $E_4(\tau)\sim (\tau')^4,
E_6(\tau)\sim (\tau')^6$, 
\be
\label{g23lim}
g_2 = \frac13 \left( \frac{\pi\tau'}{2\omega} \right)^4
\left(1+240 q'^2+\dots\right)\ ,\quad
g_3 = \frac2{27} \left( \frac{\pi\tau'}{2\omega} \right)^6
\left(1-504 q'^2+\dots\right)\ ,
\ee
provided we identify
\be
\label{rtau}
r \equiv \sqrt{x^2+4 v\bv} = \frac{\pi \tau'}{2 \I\, \omega}\ .
\ee

Using the fact that $\hat E_2(\tau)=E_2(\tau)-3/(\pi\tau_2)$ is
modular invariant of weight 2, one
obtains the asymptotic behavior of $\eta$ in \eqref{etaE2},
\be
\eta(\tau) \sim
\frac13\,\omega\,
\left(\frac{\pi \tau'}{2\omega}\right)^2\, \left( 1 - \frac{6\I}{\pi \tau'}
-24 q'^2 \right)
\sim -\frac13 \omega \, r^2 + r + \dots \ .
\ee
Thus, the K\"ahler potential reduces to
\be\label{eq:KPAH}
\begin{split}
K = 4 \, \eta - (y_+ + y_-) \, \omega
=  - 2 \, \omega \left( x^2 + 2 v \bv \right) + 4 r + \dots\ ,
\end{split}
\ee
in agreement with \eqref{TN3}, up to an overall factor of 4,
provided we set $c=\omega/2$. 

It remains to express $x,v,\bar v$
in terms of the complex coordinates $U,\bU,Z,\bZ$. Since $Z=v^2$,
the only non-trivial task is to express $x$ in terms of $U,\bU,Z,\bZ$.
Expanding $\mathcal{P}(v), \pi(v)$ on the axis $u\in\omega\IR+\omega'$
(i.e. $v\in\frac{\pi}{2}\tau+\IR$) leads to
\be
\pi(y_-) \equiv \pi(v_-) = - \omega x + {\rm arctanh}(x/r) + \dots 
\ee
which reproduces the condition \eqref{TN2} provided
one identifies $U=u, \bar U=\bu$,
 \be
 \label{UUbx}
 \frac{U+\bar U}{2} = \pi(y_-) = - \omega x +
 {\rm arcsinh}\left( x / \sqrt{4 v \bv} \right) + \dots 
 \ee
On the other hand, expanding the Weierstrass function
and the elliptic integral $\pi$ on the axis $u\in\omega+\IR\omega'$
(i.e. $v\in \frac{\pi}{2}+\I \IR$) and 
neglecting exponential corrections, the condition $\mathcal{P}(v_+)=y_+$
is trivially satisfied independently of $u_+$, and $v_+$
and related to $U-\bU$ via
\be
\frac{U-\bar U}{2} = \pi(y_+) \equiv \pi(v_+) = - \I \tau'
\left(v_+-\frac{\pi}{2}\right)\ .
\ee
The fact that the K\"ahler potential \eqref{eq:KPAH} depends only
on $Z,\bZ,U+\bU$ is of course a consequence of the triholomorphic
isometry recovered in the Taub-NUT limit. Utilizing the explicit expressions for the AH metric
obtained in \cite{Ionas:2007gd}, one can also verify that in the strict Taub-Nut-limit one exactly recovers
\eqref{TN4}, provided $U=u_{TN}, Z=v_{TN}, c=\omega/2$.

\subsubsection{Exponential deviations from the Taub-NUT limit}

Let us now incorporate the first exponential correction to the Taub-NUT limit.
Our aim is to determine the twistor line $\eta^{(4)}(\zeta)$ as a function
of the complex coordinates $U,\bar U,Z,\bar Z$, to first order in $q'$. We
thus deform the relations \eqref{zwy} into
\be
\label{zwydef}
w = 2 x v  + q' \delta w+\dots\ ,\quad
y = x^2 - 2 v \bv+ q' \delta y+\dots\ ,\quad
\ee
keeping the relation $z = v^2$ undeformed (indeed,
$Z=2\sqrt{z}$ is an exact relation, and we work at fixed $U,\bar U,Z,\bar Z$.)
Computing $y_\pm$ and $w_\pm$ from \eqref{zwydef}, inserting into \eqref{g23x}
and comparing to \eqref{g23lim}, we learn that the relation
\eqref{rtau} is deformed to
\be
\tau' = \frac{2i \omega r}{\pi} + \frac{i\omega}{\pi r^3}
\left[ \delta y ( x^2 - 2 v \bar v)+3 x \Delta \right]
q' + \dots\ ,
\ee
where $\Delta \equiv v \delta \bw+ \bv \delta w$.
Note that we define $r$ to be equal to $\sqrt{x^2+4 v\bv}$, and $x$ to
be the function of $U,\bU,Z,\bZ$ given implicitly by \eqref{UUbx}.
We also expand the uniformizing parameters $v_\pm$ associated to $y_\pm$ as
\be
v_- = -\frac{\pi}{2}\tau+\frac{\pi}{2\omega r}
{\rm arcsinh}\left(x/\sqrt{4 v\bv}\right)+q'\,\delta v_-\ ,\quad
v_+ = \frac{\pi}{4\omega r} (U-\bar U) + q' \,\delta v_+\ .
\ee
The two conditions $\mathcal{P}(v_+)=y_+$ and $\mathcal{P}(v_-)=y_-$  may
be used to compute $\delta y$ and $\delta v_-$, respectively,
\be
\delta y= \frac{8 r^4 \cosh(U-\bU) + x \Delta}{2 v \bv}\ ,
\ee
\be
\begin{split}
\delta v_- =& \frac{\pi}{8\omega v \bv r^3}
\left[ r \Delta
+ 12 x r^3 \cosh(U-\bU) \right.\\
&\left. - \left( x \Delta
+8 r^2 (x^2 - 2 v \bv )  \cosh(U-\bU) \right)
{\rm arcsinh}\left(x/\sqrt{4 v\bv}\right) \right]\ .
\end{split}
\ee
Finally, the conditions that $\pi(v_+)$ and $\pi(v_-)$ have no variation
(because $U,\bU$ are fixed) allows to compute $\delta v_+$ and $\Delta$:
\be
\Delta = - 4 r^2 x \, \frac{3-2 \omega r}{1-\omega r} \cosh(U-\bU)\ ,
\ee
\be
\delta v_+ = -2\pi \sinh(U-\bU) -\frac{\pi}{4\omega}
\frac{x^4-16 v^2 \bv^2-4 \omega r^3 v \bv + 8 r^2 v \bv}
{r^3 v \bv (1-\omega r)} (U-\bU) \cosh(U-\bU)\ .
\ee
While these relations are not sufficient to fix $\delta w$ and $\delta \bw$
separately, they are sufficient to obtain the leading exponential
correction to the K\"ahler potential \eqref{eq:KPAH},
\be
K =  - 2 \, \omega \left( x^2 + 2 v \bv \right) + 4 r -\frac{4 r^3}{v \bv}
\cosh(U-\bU)\, q' + \mathcal{O}(q'^2)\ .
\ee
Using the same spherical coordinates as for Taub-NUT space \eqref{sphtn},
this may be rewritten as
\be
K =  -\frac12 \omega r^2 ( 3 + \cos 2\theta ) + 4 r
- 16 \, \frac{\cos2\psi}{(\sin\theta)^2}\, r\, e^{-2\omega r}
+  \mathcal{O}(q'^2)\ .
\ee
The variation of $K$ may be rewritten as in  \eqref{kahlerdef} as the Penrose 
transform of the holomorphic function
\be
\label{HtAH}
H_{(1)}(\eta, \muinf, \zeta) = 4\, \eta \, \cos( 4\muinf ) \ ,
\ee
where $\eta$ and $\muinf$ are the global sections of unperturbed Taub-NUT space,
given in \eqref{exactmutnth}.
Indeed, an explicit computation shows that
\be
 \oint_{C}\,  \frac{ \de\zeta}{2\pi \I \zeta}H_{(1)}
= -16 \, \frac{\cos(2 \psi)}{(\sin\theta)^2} \,r\, e^{-2 \omega r} \,,
\ee
where $C$ denotes the figure-eight contour enclosing the two poles
$\zeta_+ = e^{\I\phi}\cot(\theta/2)$ and $\zeta_- = -e^{\I\phi}\tan(\theta/2)$ clockwise and counterclockwise, respectively. 

The main result of this section \eqref{HtAH} provides a remarkably simple and
elegant way to summarize the deviation of the AH manifold away 
from its Taub-NUT limit at leading order.  It should be noted that the exponential dependence
of $H_{(1)}$ on $\muinf$ is fixed by the requirement that instanton corrections preserve
discrete shifts $\vrh\to \vrh+ \pi$. Instanton corrections to the metric components have
been analyzed in \cite{Hanany:2000fw}. It would be interesting to extend this analysis
to other HK spaces whose metric is known only implicitly, such as the
moduli space of monopoles of higher charge~\cite{Atiyah:1988jp,Houghton:1999hr},
or  Dancer's manifold \cite{Dancer,Chalmers:1997ff}.

\acknowledgments
We are very grateful to R. Ionas for correspondence on \cite{Ionas:2007gd}, M. Ro\v{c}ek 
for informing us of his upcoming work \cite{Lindstrom:2008gs}, and to A. Neitzke for
enlightening discussions. The research of S.A. is supported by CNRS and
by the contract
ANR-05-BLAN-0029-01. The research of B.P. is supported in part by ANR(CNRS-USAR)
contract no.05-BLAN-0079-01. F.S.\ acknowledges financial support from the ANR grant BLAN06-3-137168. S.V. thanks the Federation de Recherches ``Interactions Fondamentales'' and LPTHE at Jussieu for hospitality and financial support.
Part of this work is also supported by the EU-RTN network MRTN-CT-2004-005104
``Constituents, Fundamental Forces and Symmetries of the Universe''.

\appendix

\section{Direct verification of the HK property}
\label{ap_integr}
In this appendix, we give a direct verification that the complex manifold with
\kahler potential \eqref{Kdef} is HK, to first order in the perturbation.

We first need some identities for the function $G_I$ defined in \eqref{defgi}.
Substituting the result \eqref{nupi} for $\nupi{i}^I$ and using the relations
\be
\begin{split}
{\Hpij{ij}}_I & = \p_{x^I}\Hpij{ij}+\I\Hp^{[ij]J}\(2\p_{x^I}\muzi{j}_J- (\Hij{j0}_{IJ}+\Hij{j\infty}_{IJ})\)
\\
& =\zeta\p_{v^I} \Hpij{ij}+\I\Hp^{[ij]J}\(2\zeta\p_{v^I}\muzi{j}_J-(\Hij{j0}_{IJ}+\Hij{j\infty}_{IJ})\) \, ,
\end{split}
\ee
the function \eqref{defgi} can be written in one of the two following forms
\bea
G_I &=& \p_{x^I} \Hpij{ij} - 2\I\Hp^{[ij]J}\Hij{ij}_{IJ}
\nonumber \\
&& +\I\sum_k\oint_{C_k} \frac{\de\zeta'}{2\pi \I\, \zeta'}
\( \frac{\zeta^3+\zeta'^3}{\zeta\zeta'(\zeta'-\zeta)}\, \Hp^{[ik]J}(\zeta')\Hij{ij}_{IJ}(\zeta)
+\frac{\zeta+\zeta'}{\zeta'-\zeta}\, \Hp^{[ij]J}(\zeta)\Hij{ik}_{IJ}(\zeta')\)
\nonumber \\
&=& \zeta\p_{v^I} \Hpij{ij}- 2i\Hp^{[ij]J}\Hij{ij}_{IJ}
\nonumber \\
&& +\I\sum_k\oint_{C_k} \frac{\de\zeta'}{2\pi \I\, \zeta'^2}
\( \frac{\zeta^3+\zeta'^3}{\zeta(\zeta'-\zeta)}\, \Hp^{[ik]J}(\zeta')\Hij{ij}_{IJ}(\zeta)
+\frac{\zeta(\zeta+\zeta')}{\zeta'-\zeta}\, \Hp^{[ij]J}(\zeta)\Hij{ik}_{IJ}(\zeta')\) \, ,
\nonumber
\eea
where $\zeta$ is inside of $C_j$.
These representations can be used to obtain
\bea
\oint_C \frac{\de\zeta}{2\pi \I\, \zeta}\, G_I &=& \p_{x^I}\oint_C \frac{\de\zeta}{2\pi \I\, \zeta}\, \Hp
\label{intF1}
 \\
&& -\I \(\oint_C \frac{\de\zeta}{2\pi \I\, \zeta^2}\, \Hp^J \oint_C \frac{\de\zeta}{2\pi \I}\,H_{IJ}
- \oint_C \frac{\de\zeta}{2\pi \I}\, \Hp^J \oint_C \frac{\de\zeta}{2\pi \I\, \zeta^2}\,H_{IJ} \)\, ,
\nonumber
\eea
\bea
\oint_C \frac{\de\zeta}{2\pi \I\, \zeta^2}\, G_I &=& \p_{v^I}\oint_C \frac{\de\zeta}{2\pi \I\, \zeta}\,\Hp
\label{intF2}
 \\
&& -\I \( \oint_C \frac{\de\zeta}{2\pi \I\, \zeta^2}\, \Hp^J \oint_C \frac{\de\zeta}{2\pi \I\,\zeta}\,H_{IJ}
-\oint_C \frac{\de\zeta}{2\pi \I}\, \Hp^J \oint_C \frac{\de\zeta}{2\pi \I\, \zeta^3}\,H_{IJ}\) \, ,
\nonumber
\eea
where the sum over contours and the patch indices are  suppressed. These results
may be used to check the integrability condition for the existence of the K\"ahler potential,
\be
\p_{w_I}\(\CL_{v^J}+\nupk{0}^K\CL_{v^Kv^J}+\bnupk{0}^K\CL_{\bv^Kv^J}\)
= -\p_{u^J} \(x^I+\I\CL_{\vrh_I}\)\ .
\ee

Since the Lagrangian $\CL$ is an integral of a holomorphic function of only $2d+1$ variables,
it must satisfy some relations. Indeed, it may be checked explicitly that
\bea
\CL_{x^I x^J}+\CL_{u^I \bu^J} &=&
-\I \(\CL_{x^I x^K}\CL_{\vrh_K x^J}-\CL_{x^I\vrh_K}\CL_{x^K x^J}
+\CL_{u^I\vrh_K}\CL_{x^K \bu^J}-\CL_{u^I x^K}\CL_{\vrh_K \bu^J} \)
\nonumber \\
&& -\(\CL_{x^I x^K}\CL_{x^L x^J}+\CL_{u^I x^K}\CL_{x^L \bu^J}\)\CL_{\vrh_K\vrh_L} \, ,
\label{prop1}
\\
\CL_{x^I u^J}-\CL_{u^I x^J} &=&
-\I \(\CL_{u^I x^K}\CL_{\vrh_K x^J}-\CL_{x^I\vrh_K}\CL_{x^K u^J}
+\CL_{x^I x^K}\CL_{\vrh_K u^J}-\CL_{u^I\vrh_K}\CL_{x^K x^J} \)
\nonumber \\
&& -\(\CL_{x^I x^K}\CL_{x^L u^J}-\CL_{u^I x^K}\CL_{x^L x^J}\)\CL_{\vrh_K\vrh_L} \, .
\label{prop2}
\eea
which generalize the standard identities for the unperturbed case,
\be
\CL_{(0)x^I x^J}+\CL_{(0)v^I \bv^J} =0\, ,
\qquad
\CL_{(0)x^I v^J}-\CL_{(0) v^I x^J}=0\, .
\label{relundef}
\ee
Equivalently,  \eqref{prop1} and \eqref{prop2} may be rewritten as
\bea
&& \(\p_{x^I}+\I\CL_{x^I x^K}\p_{\vrh_K}\)\(\p_{x^J}-\I\CL_{x^J x^L}\p_{\vrh_L}\)\CL
\nonumber \\
&& \qquad\qquad\qquad\qquad\qquad
+ \(\p_{u^I}-\I\CL_{u^I x^K}\p_{\vrh_K}\)\(\p_{\bu^J}+\I\CL_{\bu^J x^L}\p_{\vrh_L}\)\CL=0 \, ,
\label{prop1mod}
\\
&& \(\p_{x^I}+\I\CL_{x^I x^K}\p_{\vrh_K}\)\(\p_{u^J}-\I\CL_{u^J x^L}\p_{\vrh_L}\)\CL
\nonumber \\
&& \qquad\qquad\qquad\qquad\qquad
- \(\p_{x^J}+\I\CL_{x^J x^K}\p_{\vrh_K}\)\(\p_{u^I}-\I\CL_{u^I x^L}\p_{\vrh_L}\)\CL=0\ .
\label{prop2mod}
\eea
These properties can be used to compute the inverse metric,
\be
\begin{split}
& K^{\bu^I u^J} = -\CL^{x^I x^J}
 +\CL_{\vrh_I\vrh_J}
-\I\[\CL^{x^I x^K}\CL_{x^K\vrh_J}-\CL_{\vrh_I x^K}\CL^{x^K x^J} \]\, ,
\\
& K^{\bu^I w_J} = -\CL^{x^I x^K}\CL_{x^K u^J}
-\I\[\CL_{\vrh_I u^J}-\CL_{\vrh_I x^K}\CL^{x^K x^L}\CL_{x^L u^J}\] \, ,
\\
& K^{\bw_I u^J} = -\CL_{\bu^I x^K}\CL^{x^K x^J}
+\I\[\CL_{\bu^I \vrh_J}-\CL_{\bu^I x^M}\CL^{x^M x^N}\CL_{x^N\vrh_J}\] \, ,
\\
& K^{\bw_I w_J} = \CL_{\bu^I u^J}-\CL_{\bu^I x^K}\CL^{x^K x^L}\CL_{x^L u^J}\, .
\end{split}
\label{HKmetinv}
\ee
Recalling that  $\omega^+=\de w_I \wedge \de u^I$, it is now straightforward to check
that the complex structures satisfy the algebra of the quaternions, e.g.
\be
J^+\, J^-=-\hf\( {\bf 1}+\I J^3 \)
 \quad \leftrightarrow \quad
(\omega^+)_{\alpha\gamma}\,K^{\gamma\bar \delta}\,(\omega^-)_{\bar \delta\bar \beta}
=K_{\alpha\bar \beta} \ .
\label{relmetrics}
\ee


\providecommand{\href}[2]{#2}\begingroup\raggedright\endgroup

\end{document}